\documentclass[%
reprint,
 superscriptaddress,
 amsmath,amssymb,
 aps,
 nofootinbib,
 prb,
 physics
]{revtex4-2}

\usepackage{booktabs}                       
\usepackage{multirow}
\usepackage{bigdelim}
\usepackage[dvipdfmx]{graphicx}
\usepackage{color}
\usepackage{dcolumn}
\usepackage{physics}
\usepackage{mathtools}
\usepackage{bm}
\usepackage{array}
\usepackage{comment}
\usepackage{cancel}
\usepackage{here}
\usepackage[dvipsnames]{xcolor}
\usepackage{mathrsfs}
\usepackage{amsmath,amsthm}
\usepackage{enumerate}
\usepackage{empheq} 
\theoremstyle{definition}

\usepackage{subfig}

\def\be{\begin{equation}}
\def\ee{\end{equation}}
\def\bea{\begin{eqnarray}}
\def\eea{\end{eqnarray}}
\newcommand{\mb}[1]{\mathbb{#1}}
\newcommand{\mc}[1]{\mathcal{#1}}

\newcommand{\IQ}{\mb{Q}}
\newcommand{\IR}{\mb{R}}
\newcommand{\IZ}{\mb{Z}}

\newcommand{\pd}{\partial}

\newcommand{\re}{{\rm e}}
\newcommand{\ri}{{\mathsf{i}}}
\newcommand{\rd}{{\rm d}}

\usepackage{xcolor} 
\definecolor{RoyalBlue}{rgb}{0.255, 0.412, 0.882}
\definecolor{DeepSkyBlue}{rgb}{0.0, 0.749, 1.0}
\definecolor{Crimson}{rgb}{0.862, 0.078, 0.235}
\definecolor{ForestGreen}{rgb}{0.133, 0.545, 0.133}
\definecolor{OrangeRed}{rgb}{1.0, 0.271, 0.0}
\definecolor{Orchid}{rgb}{0.855, 0.439, 0.839}
\definecolor{Sienna}{rgb}{0.627, 0.322, 0.176}
\definecolor{Goldenrod}{rgb}{0.855, 0.647, 0.125}
\definecolor{CadetBlue}{rgb}{0.372, 0.619, 0.627}
\definecolor{CornflowerBlue}{rgb}{0.392, 0.584, 0.929}
\definecolor{RebeccaPurple}{rgb}{0.4, 0.2, 0.6}
\definecolor{Salmon}{rgb}{0.980, 0.502, 0.447}
\definecolor{HotPink}{rgb}{1.0, 0.412, 0.706}
\definecolor{Chocolate}{rgb}{0.824, 0.412, 0.118}
\definecolor{SteelBlue}{rgb}{0.275, 0.510, 0.706}
\definecolor{FireBrick}{rgb}{0.698, 0.133, 0.133}
\definecolor{bondiblue}{rgb}{0.0, 0.58, 0.71}
\definecolor{celestialblue}{rgb}{0.29, 0.59, 0.82}
\definecolor{coolblack}{rgb}{0.0, 0.18, 0.39}
\definecolor{frenchblue}{rgb}{0.0, 0.45, 0.73}
\definecolor{lapislazuli}{rgb}{0.15, 0.38, 0.61}
\definecolor{mediumpersianblue}{rgb}{0.0, 0.4, 0.65}
\definecolor{darkpowderblue}{rgb}{0.0, 0.2, 0.6}
\definecolor{darkcandyapplered}{rgb}{0.64, 0.0, 0.0}
\definecolor{darkscarlet}{rgb}{0.34, 0.01, 0.1}
\definecolor{falured}{rgb}{0.5, 0.09, 0.09}
\definecolor{darkcyan}{rgb}{0.0, 0.55, 0.55}

\def\beq{\begin{equation}}
\def\eeq{\end{equation}}
\def\bp{\begin{pmatrix}}
\def\ep{\end{pmatrix}}

\usepackage[pdftex,colorlinks,urlcolor=darkcyan,citecolor=blue,linkcolor=blue]{hyperref}
 
\graphicspath{{figs/}}

\begin{document}

\author{Jie Gu}
\affiliation{School of physics \& Shing-Tung Yau Center, Southeast University, Nanjing  211189, P. R. China}

\author{Yunfeng Jiang}
\email{jinagyf2008@seu.edu.cn}
\affiliation{School of physics \& Shing-Tung Yau Center, Southeast University, Nanjing  211189, P. R. China}
\affiliation{Peng Huanwu Center for Fundamental Theory, Hefei, Anhui 230026, China}

\author{Huajia Wang}
\email{wanghuajia@ucas.ac.cn}
\affiliation{Kavli Institute for Theoretical Sciences, University of Chinese Academy of Sciences, Beijing 100190, China}
\affiliation{Peng Huanwu Center for Fundamental Theory, Hefei, Anhui 230026, China}

\date{\today}

\title{Resurgent properties of $T\overline{T}$-deformed conformal field theories}
\begin{abstract}
We elaborate on the resurgence analysis on the $T\overline{T}$-deformed 2d conformal field theory (CFT). Writing the deformed partition function as an infinite series in the deformation parameter $\lambda$, we develop efficient analytical methods to compute high-order terms of the $\lambda$-series. Based on the asymptotic behavior of the large-order perturbative series, we show that the $\lambda$-series is asymptotic and extract non-perturabtive contributions by resurgence. This paper extends a previous Letter of the same authors to interacting rational CFTs, with Lee-Yang CFT as a concrete example. We also discuss possible implications of the resurgence results on holography.

\end{abstract}

\maketitle

\section{Introduction}
\label{sec:intro}
The $T\overline{T}$-deformation \cite{Smirnov:2016lqw,Cavaglia:2016oda} is an irrelevant deformation of 2d quantum field theories (QFT) triggered by the composite operator $T\overline{T}=\det T_{\mu\nu}$. It can be defined for any 2D QFT and has a number of remarkable properties. The deformation alters the short distance behavior of the theory significantly, leading to a new type of UV behavior which are different from the Landau pole or asymptotic safety. The deformed theory is not a local QFT in the Wilsonian sense and exhibit features similar to gravity theories. At the same time, the deformation is solvable and preserves integrability, allowing an in-depth analytical study of the deformed theory.\par

A particularly interesting class of theories under $T\overline{T}$-deformation are the 2d conformal field theories (CFTs). The infinite-dimensional Virasoro symmetry renders them highly solvable, enabling the analytical computation of many physical quantities. As a result, the deformed theories are under better analytic control than a generic deformed 2d QFT. At the same time, the AdS/CFT correspondence raises a natural question: what is the bulk dual of a $T\overline{T}$-deformed CFT? Significant progress has been made in addressing this question, with proposals including introducing a radial cutoff (cut-off AdS) \cite{McGough:2016lol,Kraus:2018xrn} for $\lambda<0$ or extending the geometry (glue-on AdS) for $\lambda>0$ \cite{Apolo:2023ckr}, as well as altering the asymptotic boundary conditions in the bulk (the mixed boundary condition proposal \cite{Guica:2019nzm}). For orbifold CFTs, a related deformation known as single-trace 
$T\overline{T}$-deformation has been introduced \cite{Giveon:2017nie,Giveon:2017myj}. In such cases, the bulk geometry is deformed into a background that interpolates between AdS spacetime and flat spacetime. These developments highlight the deep connections between 
$T\overline{T}$-deformed theories, holography, and lower dimensional quantum gravity.\par

One of the central quantities in a 2d CFT is the torus partition function, which encodes rich information about the spectrum and thermodynamic properties of the theory. An important property of the CFT torus partition function is its modular invariance. Remarkably, it has been demonstrated that the 
$T\overline{T}$-deformed theory also retains modular invariance \cite{Cardy:2018sdv,Dubovsky:2018bmo,Datta:2018thy,Aharony:2018bad}, provided the deformation parameter also transforms appropriately.  Using this property, one can derive the asymptotic density of states, which exhibits an intriguing interpolation between Cardy-like behavior in the infrared (IR) limit and Hagedorn-like behavior in the UV limit \cite{Datta:2018thy}. This transition signals the emergence of non-locality in the deformed theory, a hallmark of 
$T\overline{T}$-deformations. The modularity of the deformed partition function can be established through two complementary approaches: either by solving a flow equation that governs the torus partition function \cite{Cardy:2018sdv,Datta:2018thy} or by employing an integral representation that connects the deformed and undeformed theories \cite{Dubovsky:2018bmo,Hashimoto:2019wct}. 

The torus partition function of many CFTs can be expressed explicitly in terms of known elliptic functions, allowing for a detailed study of their analytic properties. In contrast, such simple closed-form analytic expressions for the $T\overline{T}$-deformed theories remain elusive. Instead, one can construct a formal power series expansion in the deformation parameter $\lambda$. A fundamental question regarding this power series is whether it converges. If it does, what is its radius of convergence? If not—meaning the series is asymptotic—how should the result be interpreted? Without a clear resolution to this question, further investigation into the analytic properties of the deformed partition function remains challenging.

This issue is closely tied to the broader context of non-perturbative corrections in quantum field theories, such as instantons and renormalons, which are known to play a crucial role in understanding the full structure of partition functions. In the context of $T\overline{T}$-deformation, a natural question arises: are there novel non-perturbative corrections intrinsic to the deformation? If such corrections exist, how can they be computed, and what is their interpretation in the holographic bulk dual? Addressing these questions is essential for a complete understanding of the non-perturbative structure of 
$T\overline{T}$-deformed theories and their implications for holography and quantum gravity.

A powerful mathematical tool for analyzing asymptotic series and extracting non-perturbative contributions is resurgence theory \cite{Ecalle} (see for instance \cite{Marino:2012zq,Sauzin:2014intro,Aniceto:2018bis} for introduction). It has found widespread application in theoretical physics, ranging from simple quantum mechanical systems to quantum field theories (see \cite{Aniceto:2018bis} and references therein for an overview). In recent years, there has been a renewed interest in applying resurgence theory to integrable systems, including two-dimensional integrable quantum field theories (see \cite{Serone:2024uwz} and references therein) 
and four-dimensional strongly coupled maximally supersymmetric Yang-Mills theories \cite{Bajnok:2024bqr,Bajnok:2024epf,Bajnok:2024ymr}. However, the application of resurgence techniques to $T\overline{T}$-deformed theories remains largely unexplored. To the best of our knowledge, the only instance where this approach has been employed is in the context of deformed two-dimensional Yang-Mills theory \cite{Griguolo:2022xcj,Griguolo:2022hek}, where the deformed partition function is much simpler than that of CFT. 

A significant challenge in applying resurgence theory is computing perturbative series to very high orders in order to obtain sufficient amount of perturbative data. While this task is, in principle, straightforward for $T\overline{T}$-deformed CFTs using recursion relations, a naive implementation quickly becomes prohibitively tedious in practice. To overcome this obstacle, new methods are required to streamline the process and enable the efficient derivation of high-order perturbative results. In our recent work reported in the Letter \cite{Gu:2024ogh}, we initiated the resurgence analysis of the $T\overline{T}$-deformed torus partition function by developing an efficient technique to compute higher-order terms, which we demonstrated for free theories, including both free boson and free fermion. Using this method, we successfully obtained the first 600 terms of the perturbative series. In this paper, we provide a detailed explanation of the method and discuss how to generalize it to interacting theories. The interacting theories we consider in this paper are rational conformal field theories (RCFTs) with a finite number of conformal families. We present a comprehensive example using the Lee-Yang model, accompanied by numerical results. The generalization of this method to other RCFTs is straightforward.

The high-order perturbative results for free theories, as demonstrated in our previous work \cite{Gu:2024ogh}, and for interacting models, as presented here, clearly indicate that the perturbative series is indeed asymptotic. Moreover, it exhibits a novel behavior that deviates from the conventional Gevrey-I type, suggesting a unique asymptotic behavior specific to $T\overline{T}$-deformed theories. Using these high-order results, we apply resurgence theory to numerically extract non-perturbative contributions for fixed values of the modular parameter. To gain deeper insight into the origin of these non-perturbative effects, we analyze the integral representation of the deformed partition function. Typically, non-perturbative contributions can be traced to specific saddle points of the integrand. However, we find that the naive saddle point, which considers only the integration kernel, fails to reproduce the results obtained from resurgence analysis. By incorporating additional contributions from the original undeformed partition function, we successfully match the non-perturbative contributions derived via resurgence. In this paper, we provide a detailed explanation of this mechanism, illustrating it explicitly for both free theories (free boson and free fermion) and the Lee-Yang model.

Our results shed new lights on several open questions in $T\overline{T}$-deformation. The resurgence analysis offers valuable insights into the analytical properties of the deformed partition function as a function of the deformation parameter $\lambda$, revealing a surprisingly intricate structure. It is well-known that the two different signs of $\lambda$ lead to deformed theories with distinct properties. To better understand this phenomenon, we explore the behavior of the theory as $\lambda$ is continuously deformed from one sign to the other in the complex plane. Along this path, we encounter singularities, signaling the emergence of new non-perturbative effects that must be accounted for. While our findings do not provide a complete resolution to this issue, they offer a new perspective and a promising direction for further exploration. 

In the context of single-trace $T\overline{T}$-deformation, non-perturbative completions have been proposed from both the worldsheet \cite{Dei:2024sct} and the deformed quantum field theory perspectives \cite{Benjamin:2023nts}. While our work primarily focuses on double-trace 
$T\overline{T}$-deformation, it is known that these two deformations are closely related. Specifically, by restricting to the non-twisted sector with winding number $w=1$, one obtains the double-trace  $T\overline{T}$-deformed partition function. This connection allows us to compare the non-perturbative contributions from resurgence and the proposals in \cite{Dei:2024sct,Benjamin:2023nts} in a concrete although restricted way.


The rest of this paper is structured as follows. In section~\ref{sec:TTbar}, we review key aspects and fundamental results of $T\overline{T}$-deformed 2d CFTs including the deformed spectrum and torus partiton function. In section~\ref{sec:pert} we elaborate on the efficient method for computing high-order coefficients of the $\lambda$-series, both for free theories and interacting RCFTs. In section~\ref{sec:resurg} we present detailed resurgence analysis for the $\lambda$-series for fixed modular parameters, based on the large-order perturbative results. In section~\ref{sec:holography}, we discuss possible implications of our findings on holography, both for single-trace and double-trace $T\overline{T}$-deformations. We conclude and discuss future directions in section~\ref{sec:disucss}.
We also include four appendices to explain some technical details of the resurgence theory needed in our work.

\section{$T\overline{T}$-deformation, partition function and modularity}
\label{sec:TTbar}
In this section, we review some fundamental aspects of $T\overline{T}$-deformation, with a particular focus on
$T\overline{T}$-deformed CFT.

\subsection{$T\overline{T}$-deformation}
The $T\overline{T}$-deformation of a 2d QFT, described by the Lagrangian density $\mathcal{L}^{(0)}$, is defined by
\begin{align}
\frac{\rd\mathcal{L}^{(t)}}{\rd t}=\det T_{\mu\nu}^{(t)}
\end{align}
where $\mathcal{L}^{(t)}$ is the Lagrangian density of the deformed theory, and $T_{\mu\nu}^{(t)}$ is the corresponding stress-energy tensor. At $t=0$, the undeformed theory $\mathcal{L}^{(0)}$ is recovered. It is important to note that the deformation parameter $t$ is of dimension [Length]$^2$, indicating that the deformation is irrelevant according to the classification by renormalization group theory.

The solvability of the theory comes from a special property of the operator $\det T_{\mu\nu}$. As a composite operator, it requires careful definition through point-splitting in the quantum theory \cite{Zamolodchikov:2004ce,Jiang:2019tcq}. Consider a quantum field theory (QFT) in a finite volume with circumference $R$ in 1+1 dimensions. A crucial property that enables us to solve for the finite volume spectrum is the following factorization formula for the expectation value of $\det T_{\mu\nu}$ \cite{Zamolodchikov:2004ce,Cardy:2018sdv}
\begin{align}
\label{eq:factorization}
\langle n| \det T_{\mu\nu}|n\rangle =&\,\langle n|T_{00}T_{11} - T_{01}T_{01}|n\rangle \\\nonumber
=&\,\langle n|T_{00}|n\rangle \langle n|T_{11}|n\rangle - \langle n|T_{01}|n\rangle^2,
\end{align}
where $0$ and $1$ denote the time and space components, respectively, and $|n\rangle$ represents an energy eigenstate of the Hamiltonian of the theory. The proof of this property relies solely on translational invariance and the conservation of the stress tensor, therefore the above formula holds for both the deformed and undeformed theories. 

For a Euclidean QFT, the expectation values in \eqref{eq:factorization} are related to the energy and momentum of the state in the following way:
\begin{align}
\label{eq:expectationT}
\langle n|T_{00}^{(t)}|n\rangle=&\,-\frac{\mathcal{E}_n(R,t)}{R}\,,\\\nonumber
\langle n|T_{11}^{(t)}|n\rangle=&\,-\frac{\partial \mathcal{E}_n(R,t)}{\partial R}\,,\\\nonumber
\langle n|T_{01}^{(t)}|n\rangle=&\,-\frac{\ri P_n(R)}{R}\,,\\\nonumber
\langle n|\det T_{\mu\nu}^{(t)}|n\rangle=&\,\frac{1}{R}\frac{\partial\mathcal{E}_n(R,t)}{\partial t}
\end{align}
where $\mathcal{E}_n(R,t)$ is the deformed energy of the state $|n\rangle$. The momentum $P_n(R)$ is undeformed. Substituting \eqref{eq:expectationT} into \eqref{eq:factorization}, we find:
\begin{align}
\partial_t\mathcal{E}_n(R,t)=\mathcal{E}_n(R,t)\partial_R\mathcal{E}_n(R,t)+\frac{1}{R}P_n(R)^2\,.
\end{align}
This equation is nothing but the inviscid Burgers' equation, which can be solved by the method of characteristics. For CFT, at $t=0$ we have
\begin{align}
&\mathcal{E}_n(R,0)\equiv E_n(R)=\frac{1}{R}\left(n+\bar{n}-\frac{c}{12}\right)\,,\\\nonumber 
&P_n(R)=\frac{1}{R}(n-\bar{n})\,,
\end{align}
where $n$ and $\bar{n}$ are eigenvalues of Virasoro generators $L_0$ and $\bar{L}_0$. With this initial condition, we obtain the deformed energy in a closed form
\begin{align}
\mathcal{E}_n(R,t)=\frac{R}{2t}\left(\sqrt{1+\frac{4t E_n}{R}+\frac{4t^2 P_n^2}{R^2}}-1\right)
\end{align}
For later convenience, we define the dimensionless parameter $\lambda=2t/(\pi R^2)$, in terms of which the deformed energy reads
\begin{align}
\label{eq:defEn}
\mathcal{E}_n(\lambda)=\frac{1}{\lambda\pi R}\left(\sqrt{1+2\pi\lambda R E_n+\lambda^2\pi^2 R^2P_n^2}-1\right)\,.
\end{align}

\subsection{Deformed torus partition function}
The $T\overline{T}$-deformed torus partition function is defined as
\begin{align}
\mathcal{Z}(\tau,\bar{\tau}|\lambda)=\sum_n e^{2\pi\ri \tau_1 R P_n-2\pi\tau_2 R\mathcal{E}_n(\lambda)}
\end{align}
where $\tau=\tau_1+\ri\,\tau_2$ (with $\tau_1\in\mathbb{R}$ and $\tau_2\ge 0$) is the modular parameter, and we sum over all the states in the spectrum. Here, $\mathcal{E}_n(\lambda)$ is the deformed energy given in \eqref{eq:defEn}. At $\lambda=0$, we recover the CFT torus partition function, \emph{i.e.}
\begin{align}
\mathcal{Z}(\tau,\bar{\tau}|0)=Z_{\text{CFT}}(\tau,\bar{\tau})\,.
\end{align}
The deformed partition function $\mathcal{Z}(\tau,\tau|\lambda)$ satisfies the following flow equation
\begin{align}
\label{eq:flowZ}
\partial_{\lambda}\mathcal{Z}=\left[\tau_2\partial_{\tau}\partial_{\bar{\tau}}+\frac{1}{2}\left(\partial_{\tau_2}-\frac{1}{\tau_2} \right)\lambda\partial_{\lambda}\right]\mathcal{Z}\,.
\end{align}
This result can be derived through various methods. It was initially obtained using the random geometry interpretation of the $T\overline{T}$-deformation in \cite{Cardy:2018sdv} and shortly afterward re-derived through an analysis of the series expansion in $\lambda$ \cite{Datta:2018thy,Aharony:2018bad}. 

The flow equation \eqref{eq:flowZ} resembles a diffusion-type equation. A solution can be expressed as 
\begin{align}
\label{eq:intRepZ}
    \mathcal{Z}(\tau,\bar{\tau}|\lambda)=\frac{\tau_2}{\pi\lambda}\int_{\mathcal{H}_+}\frac{\rd^2\zeta}{\zeta_2^2}
    e^{-\frac{|\zeta-\tau|^2}{\lambda\zeta_2}}Z_{\text{CFT}}(\zeta,\bar{\zeta})
\end{align}
which resembles a heat kernel solution. Here, $\mathcal{H}_+$ denotes the upper half-plane. It is straightforward to verify that the integral representation \eqref{eq:flowZ} satisfies the flow equation, as all differential operators act exclusively on the kernel  $e^{-|\zeta-\tau|^2/\lambda\zeta_2}$. This integral representation can also be derived through various approaches. Initially, it was obtained from the observation that the $T\overline{T}$-deformation of a QFT is equivalent to coupling the QFT to a two-dimensional topological gravity \cite{Dubovsky:2017cnj,Dubovsky:2018bmo}.  Later, a similar integral representation was derived for more general solvable deformations, such as the $J\bar{T}+T\bar{J}+T\overline{T}$ deformation, from a worldsheet perspective \cite{Hashimoto:2019wct}.

One of the key properties of the CFT torus partition function is its modular invariance
\begin{align}
Z_{\text{CFT}}\left(\frac{a\tau+b}{c\tau+d},\frac{a\bar{\tau}+b}{c\bar{\tau}+d}\right)=Z_{\text{CFT}}(\tau,\bar{\tau})
\end{align}
where $a,b,c,d\in\mathbb{Z}$ and satisfy $ad-bc=1$. Remarkably, modular invariance is preserved under $T\overline{T}$-deformation
\begin{align}
\label{eq:modularDefZ}
\mathcal{Z}\left(\left.\frac{a\tau+b}{c\tau+d},\frac{a\bar{\tau}+b}{c\bar{\tau}+d}\right|\frac{\lambda}{|c\tau+d|^2}\right)
=\mathcal{Z}(\tau,\bar{\tau}|\lambda)\,,
\end{align}
This follows either from the flow equation \eqref{eq:flowZ} or from the integral representation \eqref{eq:intRepZ}.

The deformed partition function $\mathcal{Z}(\tau,\bar{\tau}|\lambda)$ can be expressed as a formal power series in $\lambda$
\begin{align}
\label{eq:expandZ}
\mathcal{Z}(\tau,\bar{\tau}|\lambda)=\sum_{k=0}^{\infty}Z_k(\tau,\bar{\tau})\lambda^k\,.
\end{align}
From \eqref{eq:modularDefZ}, it follows that the coefficients $Z_k(\tau,\bar{\tau})$ has the modular property
\begin{align}
Z_k\left(\frac{a\tau+b}{c\tau+d},\frac{a\bar{\tau}+b}{c\bar{\tau}+d}\right)=(c\tau+d)^k(c\bar{\tau}+d)^kZ_k(\tau,\bar{\tau})
\end{align}
indicating that $Z_k(\tau,\bar{\tau})$ is a modular form of weight $(k,k)$. Substituting the series expansion \eqref{eq:expandZ} into the flow equation \eqref{eq:flowZ}, we obtain recursion relations
\begin{align}
\label{eq:recurs}
Z_{k+1}=\frac{1}{2}\left(\frac{2\tau_2}{k+1}D^{(k)}\overline{D}^{(k)}-\frac{k}{2\tau_2}\right)Z_k,\quad k\ge 0
\end{align}
where $D^{(k)}$ and $\overline{D}^{(k)}$ are the Ramanujan-Serre (RS) derivatives
\begin{align}
\label{eq:RSderivative}
D^{(k)}\equiv\frac{\partial}{\partial\tau}-\frac{\ri k}{2\tau_2},\quad
\overline{D}^{(k)}\equiv\frac{\partial}{\partial\bar{\tau}}+\frac{\ri k}{2\tau_2}\,.
\end{align}
The RS derivatives have the following important property: if $F(\tau,\bar{\tau})$ is a modular form of weight $(k,k')$, then $D^{(k)}F(\tau,\bar{\tau})$ and $\overline{D}^{(k')}F(\tau,\bar{\tau})$ are modular forms of weight $(k+2,k')$ and $(k,k'+2)$, respectively. Using the fact that $\tau_2^{-1}$ is a modular form of weight $(1,1)$, it is clear that if $Z_k$ is a modular form of weight $(k,k)$,  the recursion relation \eqref{eq:recurs} ensures that $Z_{k+1}$ is a modular form of weight $(k+1,k+1)$, as expected.

A fundamental question regarding the series expansion \eqref{eq:expandZ} is whether the $\lambda$-series converges for fixed values of $\tau,\bar{\tau}$. To address this question, one way is to compute $Z_k(\tau,\bar{\tau})$ explicitly. The recursion relation \eqref{eq:recurs} provides the essential tools for such computations. Starting from a known CFT partition function $Z_0(\tau,\tau)=Z_{\text{CFT}}(\tau,\bar{\tau})$, written  explicitly in terms of known elliptic functions, one can in principle calculate the coefficients $Z_k$ recursively. However, as we will demonstrate in the next section, a naive implementation of the recursion relation quickly becomes prohibitively tedious, making it difficult to obtain high-order results (\emph{e.g.}, up to a few hundred orders). Therefore, more efficient methods need to be developed.

\section{Large order perturbative methods}
\label{sec:pert}
In this section, we present an efficient method for computing large-order perturbative results, based on the recursion relation \eqref{eq:recurs} and the theory of almost holomorphic modular forms. We begin by explaining the method for free theories, where partition functions can be written explicitly in terms of known elliptic functions. Subsequently, we show how to extend this approach to interacting theories. We focus on rational conformal field theories with a finite number of conformal families.

\subsection{Free boson}
The torus partition function for the free boson is given by
\begin{align}
\label{eq:level0Boson}
Z_0^{\text{B}}(\tau,\bar{\tau})=\frac{1}{\sqrt{\tau_2}\eta(\tau)\eta(\bar{\tau})}
\end{align}
where $\eta(\tau)$ is the Dedekind $\eta$-function. It is a modular function of weight $(0,0)$. Using the recursion relation \eqref{eq:recurs}, we can compute $Z_n^{\text{B}}(\tau,\bar{\tau})$ in the $\lambda$-series. The first order result is given by
\begin{align}
Z_1^{\text{B}}(\tau,\bar{\tau})=\tau_2\,\partial_{\tau}\partial_{\bar{\tau}}Z_0^{\text{B}}(\tau,\bar{\tau})
\end{align}
Since $Z_0^{\text{B}}$ in \eqref{eq:level0Boson} depends on the Dedekind $\eta$-function, the first-order computation involves terms like $\partial_{\tau}\eta(\tau)$. Similarly, $Z_n^{\text{B}}(\tau,\bar{\tau})$ will involve higher derivatives $\partial_{\tau}^n\eta(\tau)$. However, working with such higher derivatives $\partial_{\tau}^n\eta(\tau)$ becomes increasingly cumbersome as $n$ grows. These expressions are not only inconvenient to evaluate numerically, but also lack of well-behaved modular properties, further complicating their analysis.

\paragraph{A ring of almost holomorphic forms} Fortunately, the higher derivatives of $\eta$-functions can be expressed in terms of Eisenstein series using the following relation
\begin{align}
\partial_{\tau}\eta(\tau)=\frac{\pi \ri}{12}\eta(\tau)E_2(\tau)
\end{align}
along with the Ramanujan identities 
\begin{align}
\partial_{\tau}E_2(\tau)=&\,\frac{\pi\ri}{6}\left[E_2(\tau)^2-E_4(\tau)\right],\\\nonumber
\partial_{\tau}E_4(\tau)=&\,\frac{2\pi\ri}{3}\left[E_2(\tau)E_4(\tau)-E_6(\tau)\right],\\\nonumber
\partial_{\tau}E_6(\tau)=&\,\pi\ri\left[ E_2(\tau)E_6(\tau)-E_4(\tau)^2\right].
\end{align}
Since the action of derivative $\partial_{\tau}$ on $E_2(\tau)$, $E_4(\tau)$, $E_6(\tau)$ and $\eta(\tau)$ close among themselves, it follows that all $Z_n^{\text{B}}(\tau,\bar{\tau})$ can be expressed in terms of these functions, their complex conjugates, and the non-holomorphic $\tau_2$.

It is well-known that while $E_4(\tau)$ and $E_6(\tau)$ are modular forms of weight $(4,0)$ and $(6,0)$ respecitvely, $E_2(\tau)$ is not modular. Its modularity can be restored by introducing a \emph{non-holomorphic} correction. Specifically, we define 
\begin{align}
\tilde{E}_2(\tau,\bar{\tau})=E_2(\tau)-\frac{3}{\pi\tau_2}
\end{align}
which transforms as a modular form of weight $(2,0)$, satisfying
\begin{align}
\tilde{E}_2\left(\frac{a\tau+b}{c\tau+d},\frac{a\bar{\tau}+b}{c\bar{\tau}+d}\right)=(c\tau+d)^2\tilde{E}_2(\tau,\bar{\tau})\,.
\end{align}
Given that $Z_n^{\text{B}}(\tau,\bar{\tau})$ is a modular form of weight $(n,n)$, it can always be expressed in terms of $\tilde{E}_2(\tau)$, which exhibits nice modular properties. For instance, $Z_1^{\text{B}}$ can be written as
\begin{align}
\label{eq:Z1explicit}
Z_1^{\text{B}}=2\tau_2\left[-\frac{\pi^2}{144}|\tilde{E}_2|^2+\frac{1}{8\tau_2^2}\right]Z_0^{\text{B}}
\end{align}
where 
\begin{align}
|\tilde{E}_2|^2\equiv \tilde{E}_2(\tau,\bar{\tau})\overline{\tilde{E}}_2(\tau,\bar{\tau})\,.
\end{align}
Similarly, the second-order result is given by
\begin{align}
\label{eq:Z2explicit}
Z_2^{\text{B}}=2\tau_2^2\left[\frac{\pi^4|\tilde{E}_2^2-2E_4|^2}{20736}+\frac{\pi^2|\tilde{E}_2|^2}{288\tau_2^2}-\frac{1}{32\tau_2^4} \right]Z_0^{\text{B}}\,.
\end{align}
From this, it is evident that all the $Z_n^{\text{B}}$'s are elements of the differential ring generated by
\begin{align}
\{\eta^{-1}(\tau),\tilde{E}_2(\tau,\bar{\tau}),E_4(\tau),E_6(\tau)\}
\end{align}
along with their complex conjugates.

\paragraph{RS derivatives} The natural differential operators acting on the ring are the RS derivatives $D$ and $\overline{D}$, defined in \eqref{eq:RSderivative}. From now on, we omit the upper indices for simplicity, with the understanding that the RS derivative acting on a modular form is appropriately indexed to ensure the result remains a modular form. The action of the RS derivatives on the holomorphic generators of the differential ring is given by
\begin{align}
\label{eq:DErelation}
D\eta^{-1}(\tau)=&\,-\frac{\pi\ri}{12}\eta^{-1}(\tau)\tilde{E}_2(\tau,\bar{\tau})\,,\\\nonumber
DE_4(\tau)=&\,\frac{2\pi\ri}{3}\left(\tilde{E}_2(\tau,\bar{\tau})E_4(\tau)-E_6(\tau)\right)\,,\\\nonumber
DE_6(\tau)=&\,\pi\ri\left(\tilde{E}_2(\tau,\bar{\tau})E_6(\tau)-E_4(\tau)^2\right)\,,
\end{align}
while their actions on the non-holomorphic quantities are
\begin{align}
\label{eq:DErelation2}
&D\tilde{E}_2(\tau,\bar{\tau})=\frac{\pi\ri}{6}\left(\tilde{E}_2(\tau,\bar{\tau})^2-E_4(\tau)\right),\\\nonumber
&\overline{D}\tilde{E}_2(\tau,\bar{\tau})=\frac{3\ri}{2\pi\tau_2^2}
\end{align}
and
\begin{align}
\label{eq:Dtau2}
D\tau_2=\overline{D}\tau_2=0\,.
\end{align}
In particular, the last equation indicates that when computing derivatives using the RS operators,  $\tau_2$ can be treated as a constant. In fact, we can simply set $\tau_2=1$ and only restore its presence in the end by multiplying each term of $Z_n$ by an appropriate power of $\tau_2$ so that it has the correct modular weight $(n,n)$. This significantly simplifies the calculations.

Next we observe that $Z_n^{\text{B}}$ can always be expressed as a finite sum of absolute values squared of polynomials in the ring generators, as illustrated in \eqref{eq:Z1explicit} and \eqref{eq:Z2explicit}. In order words,
\begin{align}
\label{eq:structureZ}
    Z_n^{\text{B}}=\sum_k c_{n,k} X_k\overline{X}_k=\sum_k c_{n,k}|X_k|^2,\;
\end{align}
where $c_{n,k}$ are rational functions of $\pi^2$, and
\begin{align}
\label{eq:polyRing}
X_k\in\mathbb{Z}[\tilde{E}_2(\tau,\bar{\tau}),E_4(\tau),E_6(\tau),\eta(\tau)^{-1}]
\end{align}
is of modular weight $(k,0)$ while $\overline{X}_k=X_k^*$. The action of $D\overline{D}$ on $|X_k|^2$ can be written as
\begin{align}
\label{eq:DDXX}
    D\overline{D}\left(|X_k|^2\right)&=\,|DX_k|^2+\frac{9}{4\pi^2}|\partial_{\tilde{E}_2}X_k|^2-\frac{k+4}{4}|X_k|^2\nonumber\\
&\,+|X_k-\tfrac{3}{2\pi\ri}D\partial_{\tilde{E}_2}X_k|^2-\frac{9}{4\pi^2}|D\partial_{\tilde{E}_2}X_k|^2\,.
\end{align}
Since $X_k$ is an element of the polynomial ring \eqref{eq:polyRing}, its RS derivative can be written as
\begin{align}
\label{eq:DX1}
DX_k&=\,\frac{\pi\ri}{6}(\tilde{E}_2^2-E_4)\partial_{\tilde{E}_2}X_k+\frac{\pi\ri}{12}\eta\tilde{E}_2\partial_{\eta}X_k\\\nonumber
&\,+\frac{2\pi\ri}{3}(\tilde{E}_2E_4-E_6)\partial_{E_4}X_k+\pi\ri(\tilde{E}_2E_6-E_4^2)\partial_{E_6}X_k\,.
\end{align}
Similarly,
\begin{align}
\label{eq:DX2}
D\partial_{\tilde{E}_2}X_k&=\,\frac{\pi\ri}{6}(\tilde{E}_2^2-E_4)\partial^2_{\tilde{E}_2}X_k+\frac{\pi\ri}{12}\eta\tilde{E}_2\partial_{\eta}\partial_{\tilde{E}_2}X_k\\\nonumber
&\,+\frac{2\pi\ri}{3}(\tilde{E}_2E_4-E_6)\partial_{E_4}\partial_{\tilde{E}_2}X_k\\\nonumber
&\,+\pi\ri(\tilde{E}_2E_6-E_4^2)\partial_{E_6}\partial_{\tilde{E}_2}X_k\,.
\end{align}

\paragraph{Algebraic structure} The formula \eqref{eq:DDXX} and the derivatives \eqref{eq:DX1} and \eqref{eq:DX2} may appear intricate at first glance, but they possess a nice algebraic structure that allows for significant simplifications. To uncover this structure, let us define the following operators:
\begin{align}
&D_+=\frac{1}{\pi\ri}D,\qquad D_-=-6\partial_{\tilde{E}_2}\\\nonumber
&D_0=2\tilde{E}_2\partial_{\tilde{E}_2}+4E_4\partial_{E_4}+6E_6\partial_{E_6}-\frac{1}{2}\eta^{-1}\partial_{\eta^{-1}}\,.
\end{align}
These operators form a representation of the $SL(2,\mathbb{Z})$ group, satisfying the commutation relations
\begin{align}
\label{eq:comm}
[D_+,D_-]=D_0,\qquad [D_0,D_{\pm}]=\pm 2D_{\pm}\,.
\end{align}
Next, we define the state $\chi_0$ as
\begin{align}
\chi_0(\tau)=\eta(\tau)^{-1}\,.
\end{align}
It is straightforward to verify that
\begin{align}
D_0\chi_0=-\frac{1}{2}\chi_0,\qquad D_-\chi_0=0
\end{align}
which identifies $\chi_0$ as a lowest weight state. Higher weight states are constructed by repeatedly applying $D_+$:
\begin{align}
\label{eq:defChi}
\chi_{\ell}=D_+^{\ell}\chi_0\,.
\end{align}
Each $\chi_{\ell}$ is a modular form of weight $2\ell-\tfrac{1}{2}$, satisfying
\begin{align}
\label{eq:weightD0}
D_0\chi_{\ell}=(2\ell-\tfrac{1}{2})\chi_{\ell}
\end{align}
Using the commutation relations \eqref{eq:comm} along with \eqref{eq:defChi} and \eqref{eq:weightD0}, it can be shown that
\begin{align}
D_+\chi_{\ell}=\chi_{\ell+1},\qquad D_-\chi_{\ell}=-\frac{\ell(2\ell-3)}{2}\chi_{\ell-1}\,.
\end{align}
This implies that
\begin{align}
D\partial_{\tilde{E}_2}\chi_{\ell}=-\frac{\pi\ri}{6}D_+D_-\chi_{\ell}=\frac{\ell(2\ell-3)\pi\ri}{12}\chi_{\ell}\,.
\end{align}
Consequently, if we take $X_k=\chi_{\ell}$ with $k=2\ell-\tfrac{1}{2}$, the last line of \eqref{eq:DDXX} simplifies drastically, yielding
\begin{align}
\label{eq:recurschi2}
D\overline{D}(|\chi_{\ell}|^2)=&\,\pi^2|\chi_{\ell+1}|^2-\frac{4\ell^2-2\ell-1}{8}|\chi_{\ell}|^2\\\nonumber
&\,+\frac{\ell^2(2\ell-3)^2}{64\pi^2}|\chi_{\ell-1}|^2\,.
\end{align}
For example, this formula implies that
\begin{align}
D\overline{D}(|\chi_0|^2)=\pi^2|\chi_1|^2+\frac{1}{8}|\chi_0|^2\,.
\end{align}
where
\begin{align}
\chi_1=D_+\chi_0=-\frac{1}{12}\eta^{-1}\tilde{E}_2\,.
\end{align}

\paragraph{The structure of $Z_n^{\text{B}}$} We can now begin with the initial condition given in \eqref{eq:level0Boson} and apply the recursion relation \eqref{eq:recurs}. By treating $\tau_2$ as a constant and in fact setting it to 1, the repeated action of the RS derivatives can be efficiently computed using the simplified formula \eqref{eq:recurschi2}. This approach allows us to express $Z_n^{\text{B}}$ in the form
\begin{align}
\label{eq:Znchil}
Z_n^{\text{B}}=\sum_{\ell=0}^nc'_{n,\ell}|\chi_{\ell}|^2,\;
\end{align}
where the almost holomorphic components $\chi_{\ell}$ are constructed using \eqref{eq:defChi}, and the coefficients 
$c'_{n,\ell}$ are determined through the recursion relations \eqref{eq:recurschi2}. We restore the presence of $\tau_2$ in the end by multiplying each term by an appropriate power of $\tau_2$ to make sure that $Z^{\text{B}}_n$ has modular weight $(n,n)$. This systematic method streamlines the computation of higher-order terms.

\subsection{Free fermion}
The structure described above can be extended to the case of the free fermion. The partition function for the free fermion is given by
\begin{align}
Z_0^{\text{F}}(\tau,\bar{\tau})=Z_0^{(\text{A,P})}+Z_0^{(\text{A,A})}+Z_0^{(\text{P,A})}
\end{align}
where the superscripts denote different spin structures. The terms on the right-hand side are explicitly expressed as
\begin{align}
Z_0^{(\text{A,P})}(\tau,\bar{\tau})=&\,\sqrt{\frac{\theta_2(\tau)\theta_2(\bar{\tau})}{\eta(\tau)\eta(\bar{\tau})}}\,,\\\nonumber
Z_0^{(\text{A,A})}(\tau,\bar{\tau})=&\,\sqrt{\frac{\theta_3(\tau)\theta_3(\bar{\tau})}{\eta(\tau)\eta(\bar{\tau})}}\,,\\\nonumber
Z_0^{(\text{P,A})}(\tau,\bar{\tau})=&\,\sqrt{\frac{\theta_4(\tau)\theta_4(\bar{\tau})}{\eta(\tau)\eta(\bar{\tau})}}\,.
\end{align}
where $\theta_a(\tau)$, $a=2,3,4$ are the Jacobi $\theta$-functions. The partition functions for different spin structures are interrelated through modular transformations. Specifically, the partition function for the (A,P) spin structure can be mapped to those for the (P,A) and (A,A) spin structures via an S-transformation and a TS-transformation, respectively. Consequently, in computing higher-order partition functions, it suffices to focus on the (A,P) contribution, as the results for the other spin structures can be derived by applying the appropriate modular transformations. 

\paragraph{A ring of almost holomorphic forms} From now on, we focus on $Z_0^{\text{(A,P)}}$. Similar to the bosonic case, $Z_n^{\text{F}}$ will involve higher order derivatives $\partial_{\tau}^n\eta(\tau)$ and $\partial_{\tau}^n\theta_a(\tau)$. These higher-order derivatives can again be rewritten in terms of known functions, such as the $\theta$-functions and the Eisenstein series. Specifically, in addition to the relations \eqref{eq:DErelation}, \eqref{eq:DErelation2}, and \eqref{eq:Dtau2}, we have the following additional identities:
\begin{align}
D\theta_2=&\,\frac{\pi\ri}{12}\theta_2\left(\tilde{E}_2+\theta_3^4+\theta_4^4\right)\,,\\\nonumber
D\theta_3=&\,\frac{\pi\ri}{12}\theta_3\left(\tilde{E}_2+\theta_2^4-\theta_4^4\right)\,,\\\nonumber
D\theta_4=&\,\frac{\pi\ri}{12}\theta_4\left(\tilde{E}_2-\theta_2^4-\theta_3^4\right)\,.
\end{align}
Thus, the almost holomorphic modular forms for the free fermion theory can be generated by $\theta_a$, $\eta$, $E_4$, $E_6$ and $\tilde{E}_2$. In practice, however, we observe that $\eta$-function and $\theta$-functions always appear in the following two combinations
\begin{align}
\hat{\theta}_2=\sqrt{\theta_2/\eta},\qquad \Theta_{34}=\theta_3^4+\theta_4^4\,.
\end{align}
Here, $\hat{\theta}_2$ and $\Theta_{34}$ are modular forms of weight $(0,0)$ and $(2,0)$ respectively. It is straightforward to verify that they satisfy
\begin{align}
D\hat{\theta}_2=&\,\frac{\pi\ri}{24}\hat{\theta}_2\Theta_{34}\,,\\\nonumber
D\Theta_{34}=&\,\frac{\pi\ri}{3}(\tilde{E}_2\Theta_{34}-\Theta_{34}^2+2E_4)\,.
\end{align}
To conclude, for free fermion, the differential ring is generated by
\begin{align}
\{\hat{\theta}_2,\Theta_{34},\tilde{E}_2,E_4,E_6\}
\end{align}
and their complex conjugates.

\paragraph{RS derivative and algebraic structure} In the fermionic case, $Z_n^{\text{F}}$ also take the form
\eqref{eq:structureZ}, but with
\begin{align}
X_k\in\mathbb{Z}[\hat{\theta}_2,\Theta_{34},\tilde{E}_2,E_4,E_6]\,.
\end{align}
We need to compute $D\overline{D}(|X_k|^2)$ in the new polynomial ring. The procedure is the same as in the bosonic case and exhibit an $SL(2,\mathbb{Z})$ algebraic structure. More precisely, define
\begin{align}
&D_+=\frac{1}{\pi\ri}D,\qquad D_-=-6\partial_{\tilde{E}_2}\,,\\\nonumber
&D_0=2\tilde{E}_2\partial_{\tilde{E}_2}+4E_4\partial_{E_4}+6 E_6\partial_{E_6}+2\Theta_{34}\partial_{\Theta_{34}}
\end{align}
where the RS derivative $D$ in the ring takes the following form
\begin{align}
D=&\,\frac{\pi\ri}{24}\hat{\theta}_2\Theta_{34}\partial_{\hat{\theta}_2}+\frac{\pi\ri}{3}(\tilde{E}_2\Theta_{34}-\Theta_{34}^2+2E_4)\partial_{\Theta_{34}}\\\nonumber
&\,+\frac{\pi\ri}{6}(\tilde{E}_2^2-E_4)\partial_{\tilde{E}_2}+\frac{2\pi\ri}{3}(\tilde{E}_2E_4-E_6)\partial_{E_4}\\\nonumber
&\,+\pi\ri(\tilde{E}_2E_6-E_4^2)\partial_{E_6}\,.
\end{align}
We can verify that $D_{\pm}$ and $D_0$ defined above again satisfy the algebraic relation \eqref{eq:comm}. Similar to the bosonic case, we now take the lowest weight state to be
\begin{align}
\chi_0=\hat{\theta}_2,
\end{align}
satisfying
\begin{equation}
    D_0\chi_0 = 0,\qquad D_-\chi_0 = 0.
\end{equation}
Higher weight states are constructed as
\begin{equation}
    \chi_\ell = D_+^\ell \chi_0.
\end{equation}
Each $\chi_\ell$ is a modular form of weight $(2\ell,0)$, satisfying
\begin{equation}
    D_0\chi_\ell = 2\ell\chi_\ell.
\end{equation}
Similarly, it can be shown that
\begin{equation}
    D_+\chi_\ell = \chi_{\ell+1},\qquad 
    D_-\chi_\ell = -\ell(\ell-1)\chi_{\ell-1},
\end{equation}
which imply that
\begin{equation}
    D\pd_{\tilde{E}_2}\chi_\ell = -\frac{\pi\ri}{6}D_+D_-\chi_\ell = \frac{\ell(\ell-1)\pi\ri}{6}\chi_\ell.
\end{equation}
Consequently, if we take $X_k=\chi_\ell$ with $k=2\ell$, the last line of \eqref{eq:DDXX} simplifies to
\begin{equation}
\label{eq:DD-fermion}
    D\overline{D}(|\chi_\ell|^2) = \pi^2|\chi_{\ell+1}|^2 -\frac{\ell^2}{2}|\chi_\ell|^2 + \frac{\ell^2(\ell-1)^2}{16\pi^2}|\chi_{\ell-1}|^2.
\end{equation}
Then the procedure of calculation is exactly the same as the free boson.

\subsection{Rational CFTs}

An important feature of RCFTs is that they have a finite number of characters, and the torus partition function is the linear combination of their modulus squares. On the other hand, if there are $n<\infty$ independent characters in the theory, they can be regarded as the $n$-independent solutions of an $n$-th order linear differential equation, which should furthermore be modular invariant. Such a differential equation is called a Modular Linear Differential Equation (MLDE), which was first developed in \cite{Mathur:1988na} as a means to classify RCFTs (see \cite{Mukhi:2019xjy} for recent development).

For our purpose, the existence of MLDEs means that characters of an RCFT and their RS derivatives also form a differential ring with \emph{finite} generators, which makes it possible to apply the same method developed above to streamline the calculation of the $T\overline{T}$ coefficients $Z_n$ of the theory.

We will use the Lee-Yang model to illustrate our method.
The Lee-Yang model is the minimal model $\mc{M}(5,2)$. It has two primary fields
\begin{equation}
  \phi_{(1,1)}=\mb{I},\quad\phi_{(1,2)}=\phi_{(1,3)}=\Phi.
\end{equation}
Correspondingly there are two holomorphic characters $\chi_{(1,1)}$,
$\chi_{(1,2)}$.  The torus partition function is 
\begin{equation}
  Z_{(5,2)} = |\chi_{(1,1)}|^2+|\chi_{(1,2)}|^2.
\end{equation}

We need to calculate the $T\overline{T}$ coefficients of this partition
function.  As in the case of free fermion, we can consider the two
sectors $(1,1)$ and $(1,2)$ separately, as the recursion relations
that produce the $T\overline{T}$ coefficients are linear.  Furthermore, the two
sectors are related to each other by S-transformation, so we only need
to consider the calculation of the $T\overline{T}$ coefficients in one sector,
say the $(1,1)$ sector.  In particular, we should focus on the RS
derivatives acting on $\chi_{(1,1)}$, and the differential ring that they produce.

\paragraph{A ring of almost holomorphic forms} 
For this purpose, it is very useful to recall the modular linear
differential equations (MLDE) \cite{Mathur:1988na} satisfied by the
characters.  The two characters are solutions to the 2nd order ODE
\begin{equation}
  \pd_\tau^2 \chi - \frac{\pi\ri}{3} E_2 \pd_\tau \chi +\mu
  \pi^2E_4\chi = 0, 
\end{equation}
where $\mu = 11/900$.
In terms of Ramanujan-Serre derivative, the MLDE reads
\begin{equation}
  D^2\chi -\frac{\pi\ri}{3}\tilde{E}_2 D\chi + \mu\pi^2 E_4\chi = 0.
\end{equation}
It is equivalent to 
\begin{equation}
  DX + A X = 0
\end{equation}
where
\begin{equation}
  A =
  \begin{pmatrix}
    0 & -1 \\ \mu\pi^2 E_4 & -\frac{\pi\ri}{3}\tilde{E}_2
  \end{pmatrix},\quad
  X =
  \begin{pmatrix}
    \chi \\ D\chi
  \end{pmatrix}.
\end{equation}
Thus if we define
\begin{equation}
  S:=\chi_{(1,1)}, \quad T:=\frac{1}{\pi\ri}DS = \frac{1}{\pi\ri}\pd_\tau S,
\end{equation}
which are modular forms of weights $(0,0)$ and $(2,0)$ respectively, they satisfy
\begin{align}
    DS = &\pi\ri\,T,\\\nonumber
    DT = &\pi\ri\,(\frac{1}{3}\tilde{E}_2T+\mu E_4 S).
\end{align}
It is obvious then that for the $(1,1)$ sector of the Lee-Yang model, the differential ring is generated by
\begin{equation}
    \{S(\tau),T(\tau),\tilde{E}_2(\tau,\bar{\tau}),E_4(\tau),E_6(\tau)\},
\end{equation}
along with their complex conjugates.

\paragraph{RS derivative and algebraic structure}
In the Lee-Yang model, the higher order partition functions also take the form \eqref{eq:structureZ}, but with
\begin{equation}
    X_k \in R_{\text{LY}}=\IZ[\tilde{E}_2(\tau,\bar{\tau}),E_4(\tau),E_6(\tau),S(\tau),T(\tau)].
\end{equation}
We need to compute $D\overline{D}(|X_k|^2)$ in the new polynomial ring. The procedure is the same as in the free theories and exhibit an $SL(2,\mathbb{Z})$ algebraic structure. More precisely, define
\begin{align}
&D_+=\frac{1}{\pi\ri}D,\qquad D_-=-6\partial_{\tilde{E}_2}\,,\\\nonumber
&D_0=2T\pd_T+2\tilde{E}_2\pd_{\tilde{E}_2}+ 4E_4\pd_{E_4} + 6E_6\pd_{E_6},
\end{align}
where the RS derivative $D$ in the ring takes the following form
\begin{align}
D=&T\pd_S + (\frac{1}{3}\tilde{E}_2 T + \mu E_4
      S)\pd_T+\frac{1}{6}(\tilde{E}_2^2-E_4)\pd_{\tilde{E}_2}\\\nonumber
&+\frac{2}{3}(\tilde{E}_2E_4-E_6)\pd_{E_4} +
      (\tilde{E}_2E_6-E_4^2)\pd_{E_6}.  
\end{align}
We can verify that $D_{\pm}$ and $D_0$ defined above again satisfy the algebraic relation \eqref{eq:comm}. Similar to the free theories, we now take the lowest weight state to be
\begin{align}
\chi_0=S,
\end{align}
satisfying
\begin{equation}
    D_0\chi_0 = 0,\qquad D_-\chi_0 = 0.
\end{equation}
Higher weight states are constructed as
\begin{equation}
    \chi_\ell = D_+^\ell \chi_0.
\end{equation}
Each $\chi_\ell$ is a modular form of weight $(2\ell,0)$, satisfying
\begin{equation}
    D_0\chi_\ell = 2\ell\chi_\ell.
\end{equation}
Similarly, it can be shown that
\begin{equation}
    D_+\chi_\ell = \chi_{\ell+1},\qquad 
    D_-\chi_\ell = -\ell(\ell-1)\chi_{\ell-1},
\end{equation}
which imply that
\begin{equation}
    D\pd_{\tilde{E}_2}\chi_\ell = -\frac{\pi\ri}{6}D_+D_-\chi_\ell = \frac{\ell(\ell-1)\pi\ri}{6}\chi_\ell.
\end{equation}
Consequently, if we take $X_k=\chi_\ell$ with $k=2\ell$, the last line of \eqref{eq:DDXX} simplifies to
\begin{equation}
\label{eq:DD-LY}
    D\overline{D}(|\chi_\ell|^2) = \pi^2|\chi_{\ell+1}|^2 -\frac{\ell^2}{2}|\chi_\ell|^2 + \frac{\ell^2(\ell-1)^2}{16\pi^2}|\chi_{\ell-1}|^2.
\end{equation}
Then the procedure of calculation is exactly the same as in the example of free theories.

\section{Resurgence analysis}
\label{sec:resurg}

\subsection{Resurgent properties of perturbation series}
\label{sec:Large-order}

\begin{figure}
  \centering
  \subfloat[$s_n$]{\includegraphics[width=0.7\linewidth]{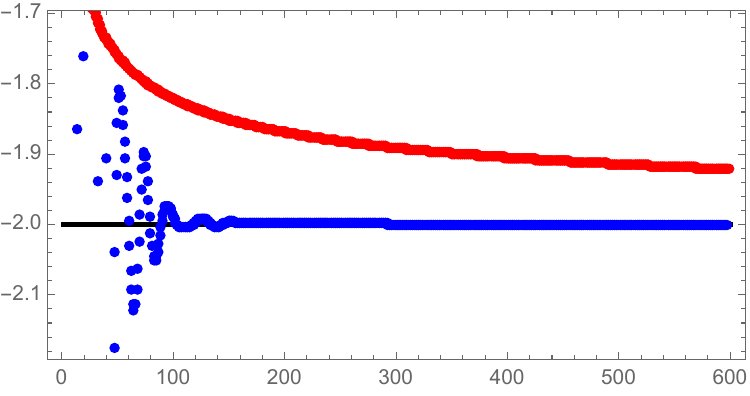}}\\
  \subfloat[$s'_n$]{\includegraphics[width=0.7\linewidth]{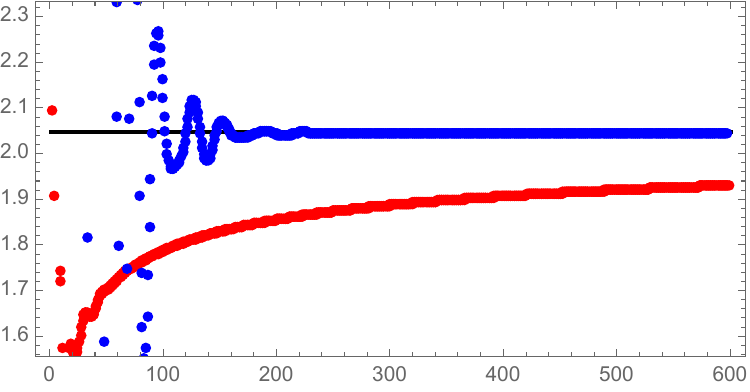}}\\
  \subfloat[$s''_n$]{\includegraphics[width=0.7\linewidth]{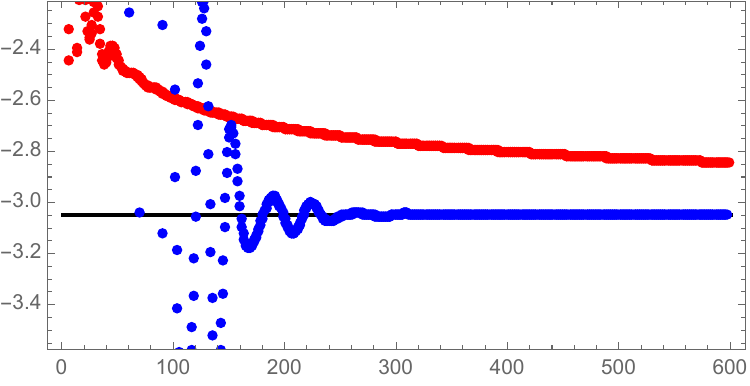}}
  \caption{Comparing $s_n$ (red dots, resp.~$s'_n$ and $s''_n$) and
    their Richardson transforms of order-$1/2$ (blue dots) against
    their asymptotic values (black line), for the example of
    $T\overline{T}$-deformed free boson at $\tau=\ri$.}
  \label{fig:s3bt1}
\end{figure}

\begin{figure}
  \centering
  \subfloat[$s_n$]{\includegraphics[width=0.7\linewidth]{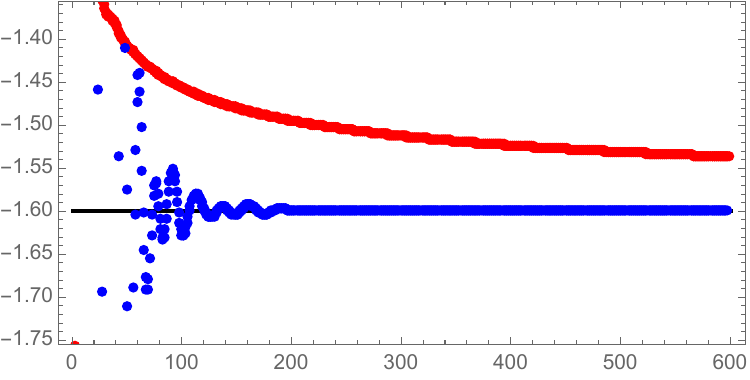}}\\
  \subfloat[$s'_n$]{\includegraphics[width=0.7\linewidth]{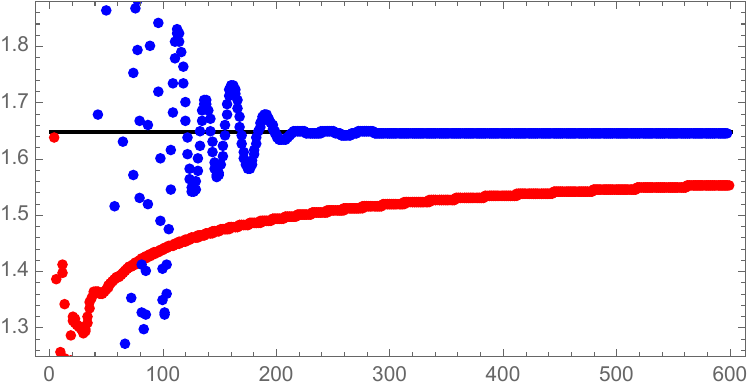}}\\
  \subfloat[$s''_n$]{\includegraphics[width=0.7\linewidth]{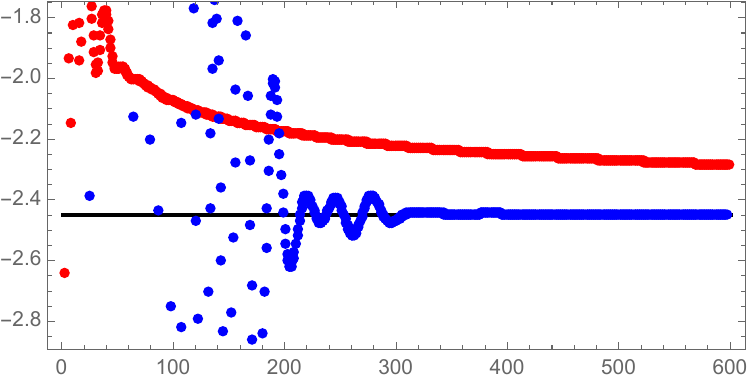}}
  \caption{Comparing $s_n$ (red dots, resp.~$s'_n$ and $s''_n$) and
    their Richardson transforms of order-$1/2$ (blue dots) against
    their asymptotic values (black line), for the example of
    $T\overline{T}$-deformed free boson at $\tau=4/5\ri$.}
  \label{fig:s3bt4o5}
\end{figure}

\begin{figure}
  \centering
  \subfloat[$s_n$]{\includegraphics[width=0.7\linewidth]{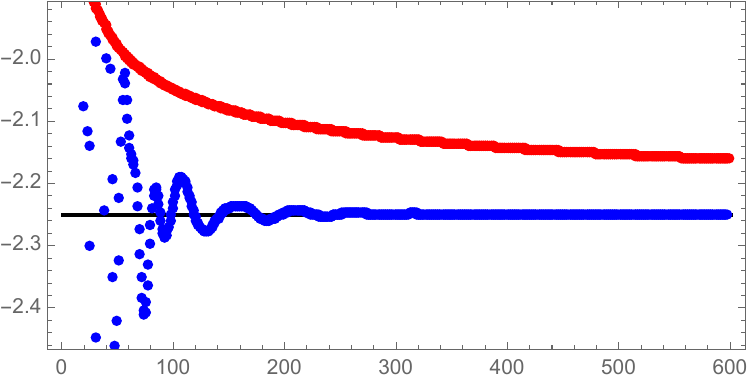}}\\
  \subfloat[$s'_n$]{\includegraphics[width=0.7\linewidth]{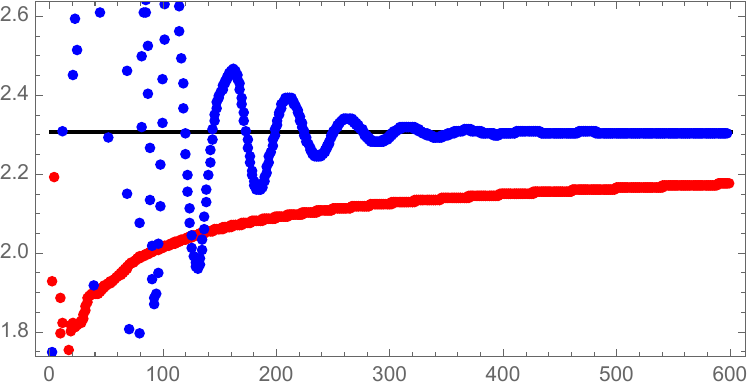}}\\
  \subfloat[$s''_n$]{\includegraphics[width=0.7\linewidth]{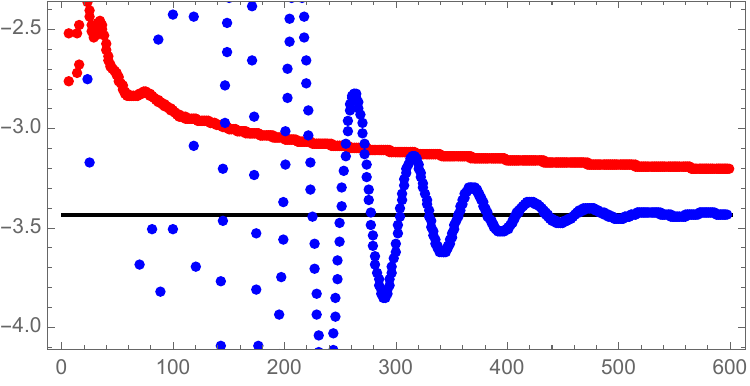}}
  \caption{Comparing $s_n$ (red dots, resp.~$s'_n$ and $s''_n$) and
    their Richardson transforms of order-$1/2$ (blue dots) against
    their asymptotic values (black line), for the example of
    $T\overline{T}$-deformed free boson at $\tau=9/8\ri$.}
  \label{fig:s3bt9o8}
\end{figure}

\begin{figure}
  \centering
  \subfloat[$s_n$]{\includegraphics[width=0.7\linewidth]{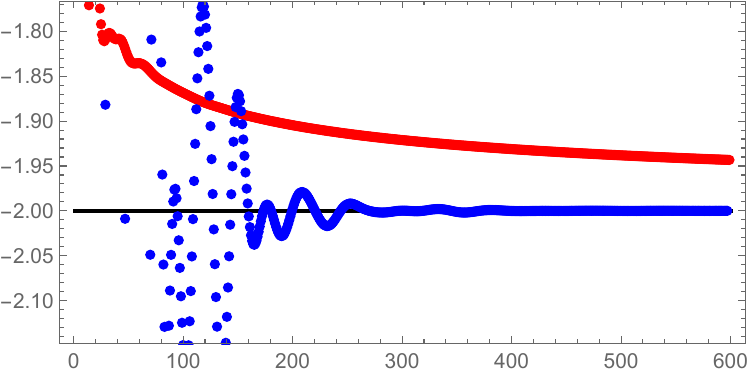}}\\
  \subfloat[$s'_n$]{\includegraphics[width=0.7\linewidth]{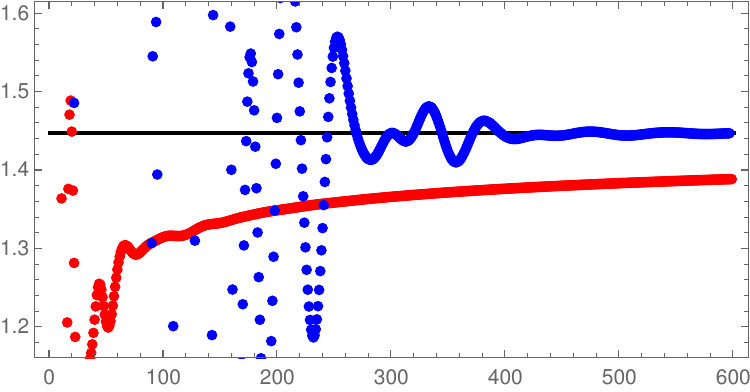}}\\
  \subfloat[$s''_n$]{\includegraphics[width=0.7\linewidth]{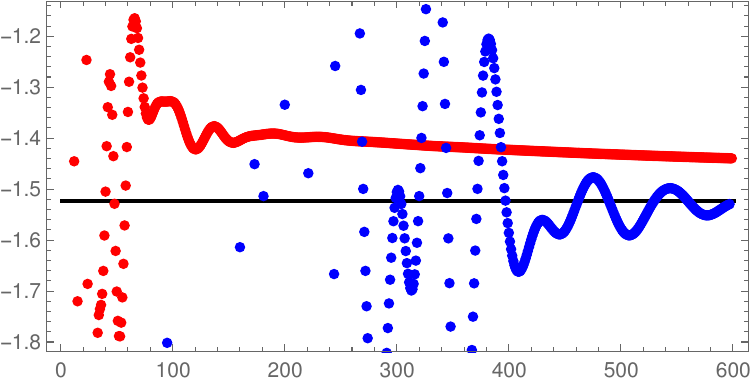}}
  \caption{Comparing $s_n$ (red dots, resp.~$s'_n$ and $s''_n$) and
    their Richardson transforms of order-$1/2$ (blue dots) against
    their asymptotic values (black line) in the example of
    $T\overline{T}$-deformed free fermion at $\tau=\ri$.}
  \label{fig:s3ft1}
\end{figure}

\begin{figure}
  \centering
  \subfloat[$s_n$]{\includegraphics[width=0.7\linewidth]{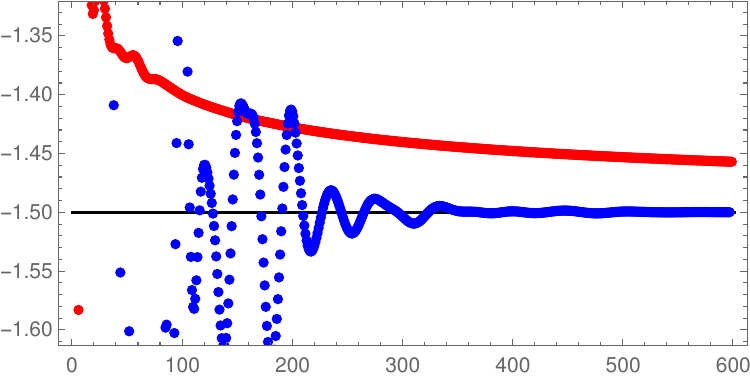}}\\
  \subfloat[$s'_n$]{\includegraphics[width=0.7\linewidth]{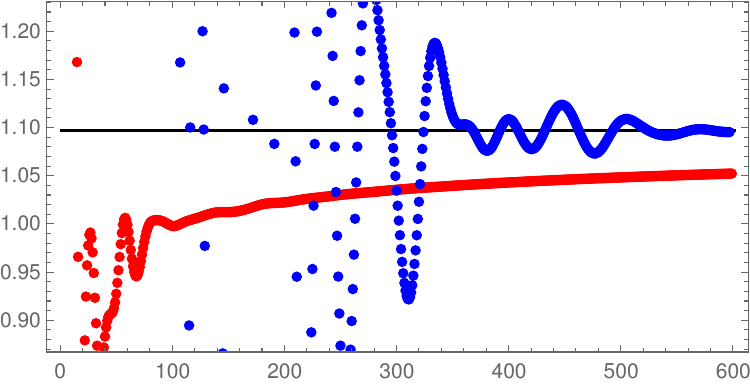}}\\
  \subfloat[$s''_n$]{\includegraphics[width=0.7\linewidth]{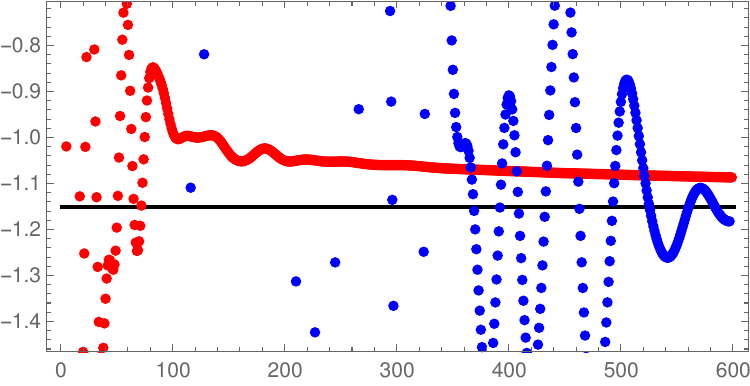}}
  \caption{Comparing $s_n$ (red dots, resp.~$s'_n$ and $s''_n$) and
    their Richardson transforms of order-$1/2$ (blue dots) against
    their asymptotic values (black line), for the example of
    $T\overline{T}$-deformed free fermion at $\tau=3/4\ri$.}
  \label{fig:s3ft3o4}
\end{figure}

\begin{figure}
  \centering
  \subfloat[$s_n$]{\includegraphics[width=0.7\linewidth]{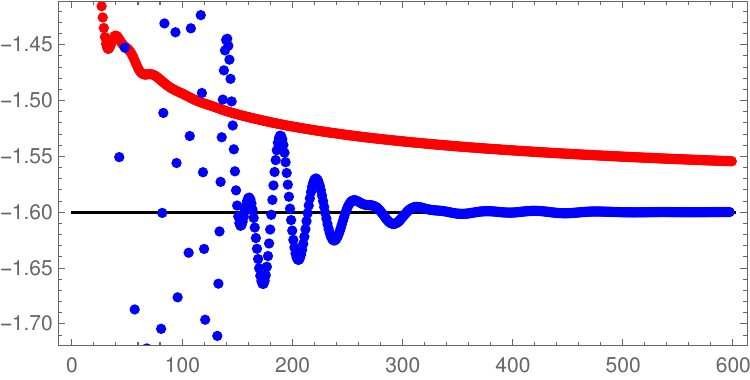}}\\
  \subfloat[$s'_n$]{\includegraphics[width=0.7\linewidth]{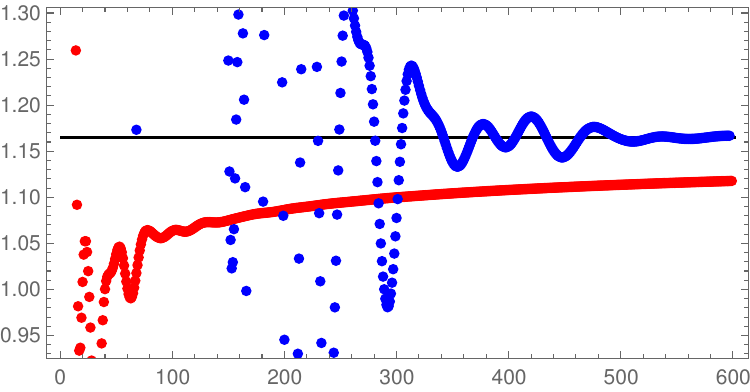}}\\
  \subfloat[$s''_n$]{\includegraphics[width=0.7\linewidth]{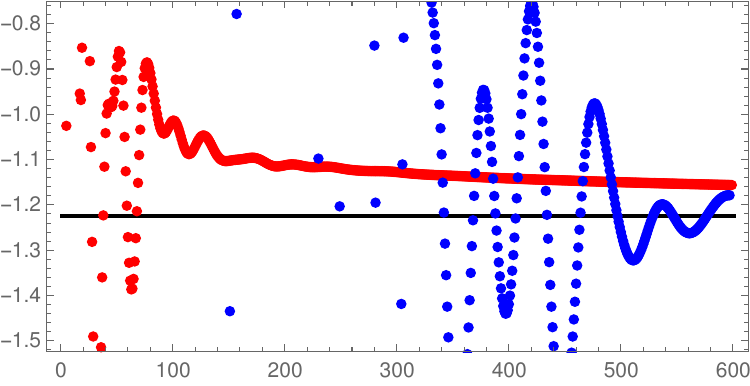}}
  \caption{Comparing $s_n$ (red dots, resp.~$s'_n$ and $s''_n$) and
    their Richardson transforms of order-$1/2$ (blue dots) against
    their asymptotic values (black line), for the example of
    $T\overline{T}$-deformed free fermion at $\tau=4/5\ri$.}
  \label{fig:s3ft4o5}
\end{figure}

\begin{figure}
  \centering
  \subfloat[$s_n$]{\includegraphics[width=0.7\linewidth]{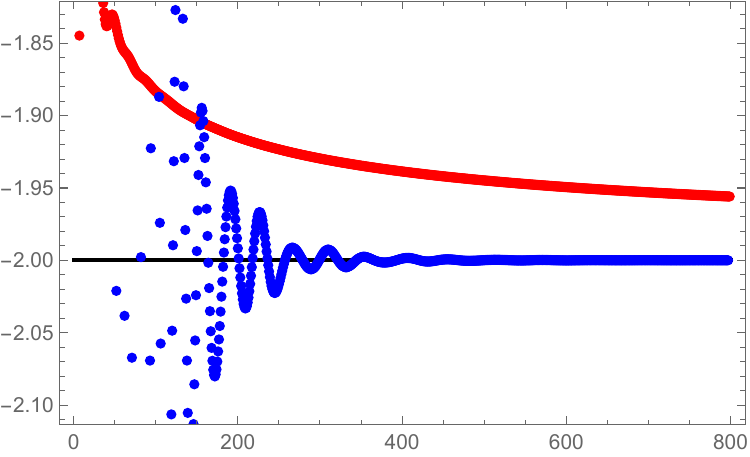}}\\
  \subfloat[$s'_n$]{\includegraphics[width=0.7\linewidth]{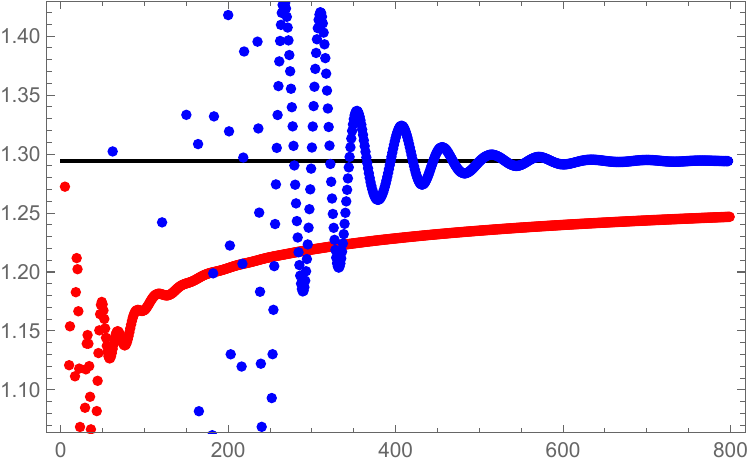}}\\
  \subfloat[$s''_n$]{\includegraphics[width=0.7\linewidth]{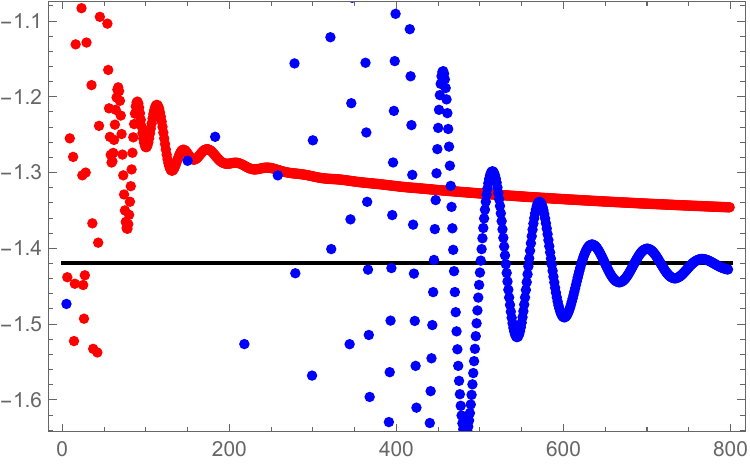}}
  \caption{Comparing $s_n$ (red dots, resp.~$s'_n$ and $s''_n$) and
    their Richardson transforms of order-$1/2$ (blue dots) against
    their asymptotic values (black line) in the example of
    $T\overline{T}$-deformed Lee-Yang model at $\tau=\ri$.}
  \label{fig:s3LYt1}
\end{figure}

\begin{figure}
  \centering
  \subfloat[$s_n$]{\includegraphics[width=0.7\linewidth]{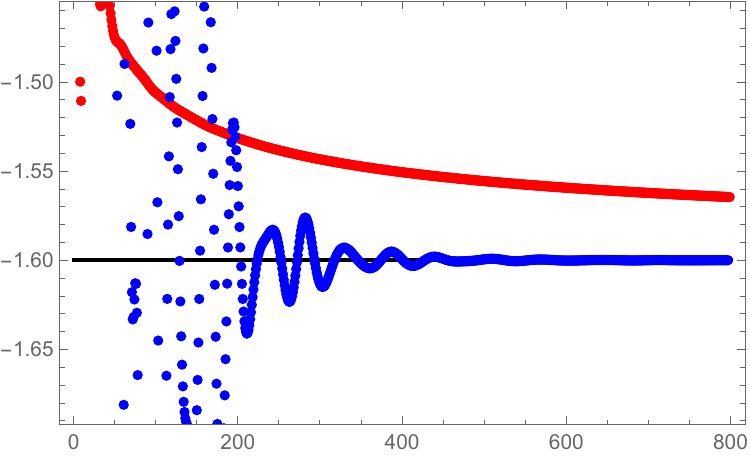}}\\
  \subfloat[$s'_n$]{\includegraphics[width=0.7\linewidth]{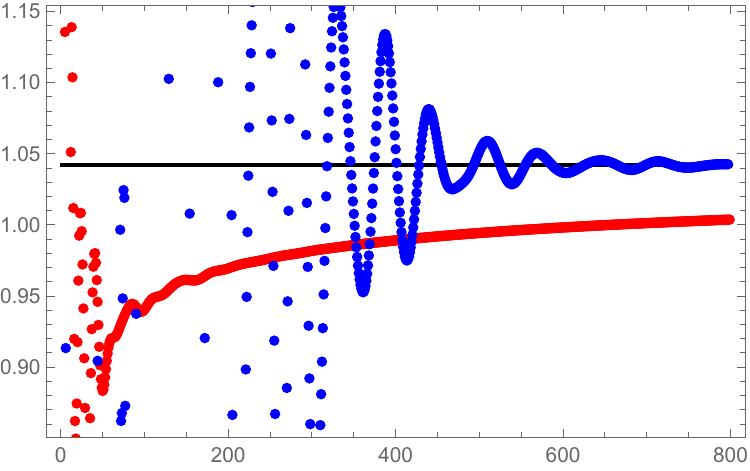}}\\
  \subfloat[$s''_n$]{\includegraphics[width=0.7\linewidth]{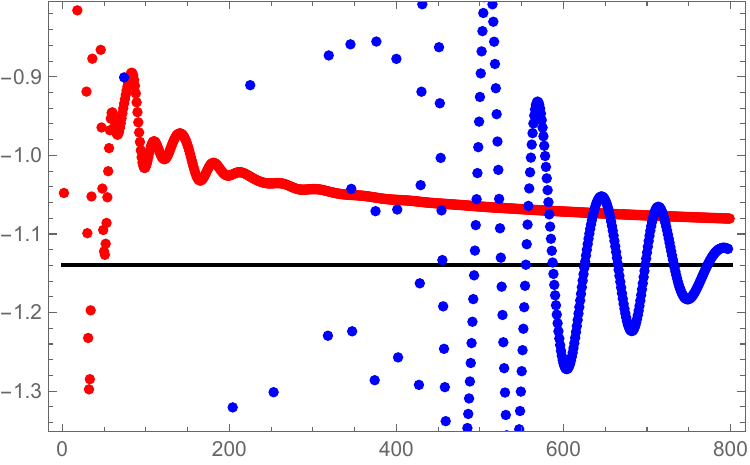}}
  \caption{Comparing $s_n$ (red dots, resp.~$s'_n$ and $s''_n$) and
    their Richardson transforms of order-$1/2$ (blue dots) against
    their asymptotic values (black line), for the example of
    $T\overline{T}$-deformed Lee-Yang model at $\tau=4/5\ri$.}
  \label{fig:s3LYt4o5}
\end{figure}

\begin{figure}
  \centering
  \subfloat[$s_n$]{\includegraphics[width=0.7\linewidth]{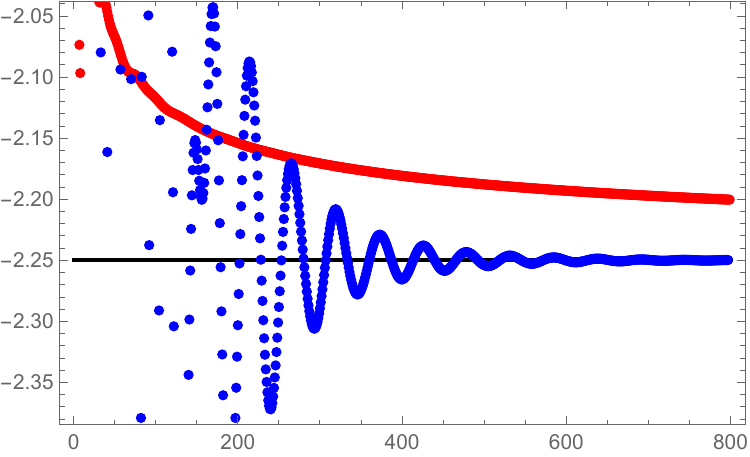}}\\
  \subfloat[$s'_n$]{\includegraphics[width=0.7\linewidth]{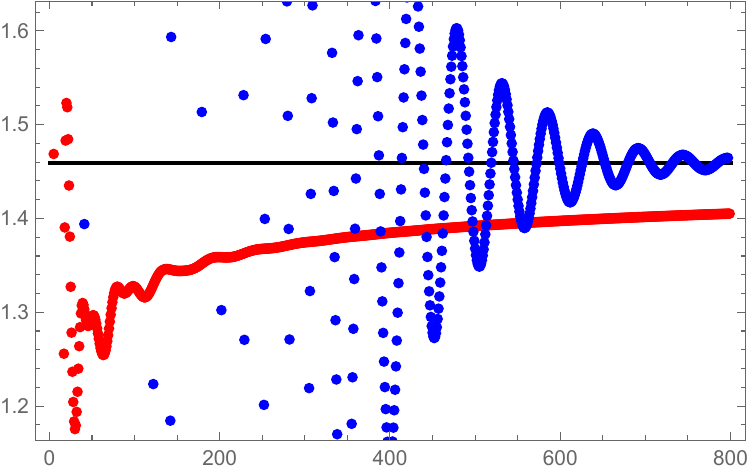}}\\
  \subfloat[$s''_n$]{\includegraphics[width=0.7\linewidth]{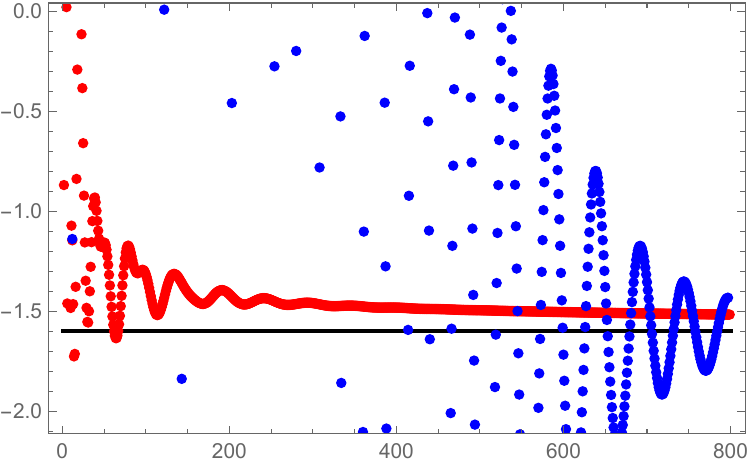}}
  \caption{Comparing $s_n$ (red dots, resp.~$s'_n$ and $s''_n$) and
    their Richardson transforms of order-$1/2$ (blue dots) against
    their asymptotic values (black line), for the example of
    $T\overline{T}$-deformed Lee-Yang model at $\tau=9/8\ri$.}
  \label{fig:s3LYt9o8}
\end{figure}

Using the method explained in Section~\ref{sec:pert}, we calculated the coefficients $Z_n$  up to $n=600$ for both the $T\overline{T}$-deformed free boson and free fermion, as well as $n=800$ coefficients $Z_n$ for $T\overline{T}$-deformed Lee-Yang model, and we evaluated these coefficients for various values of $\tau$.
We find that the $\lambda$-series of the $T\overline{T}$-deformed partition function is indeed an asymptotic series, in the sense that the coefficients $Z_n$ grow in a factorial manner.
On the other hand, these coefficients display a slightly unusual asymptotic behavior
\begin{equation}
\label{eq:Zn-asymp}
    Z_n \sim \frac{\Gamma(n+\nu)}{A^{n+\nu}}\exp\left(B\sqrt{\frac{n+\nu}{A}}\right).
\end{equation}
This asymptotic behavior can be verified numerically.
For instance, \eqref{eq:Zn-asymp} indicates the asymptotic behavior
\begin{equation}
\label{eq:resurgence_prediction}
    nZ_n/Z_{n+1} \sim A -\frac{B}{2}\sqrt{\frac{A}{n}} + \frac{B^2-8A\nu}{8n} +\ldots.
\end{equation}
Then the auxiliary sequences
\begin{subequations}
\begin{align}
    &s_n := n Z_n/Z_{n+1},\\
    &s'_n := \sqrt{n}(s_n-A),\\
    &s''_n := n(s_n-A-(B/2)\sqrt{A/n}),
\end{align}
\end{subequations}
have the following asymptotic behavior in the large $n$ limit
\begin{subequations}
\begin{align}
    s_n &\sim A + O\left(1/\sqrt{n}\right),\label{eq:s1asymp}\\
    s'_n &\sim -\frac{1}{2}B\sqrt{A}+ O(1/\sqrt{n}),\label{eq:s2asymp}\\
    s''_n &\sim \frac{1}{8}(B^2-8A\nu)+ O(1/\sqrt{n}).\label{eq:s3asymp}
\end{align}
\end{subequations}
We find numerically that indeed the sequences $s_n,s'_n,s''_n$ behave asymptotically as $\text{const.}+O(1/\sqrt{n})$ in all three models, and by comparing with \eqref{eq:s1asymp},\eqref{eq:s2asymp},\eqref{eq:s3asymp}, the parameters are read off to be
\begin{equation}
\label{eq:pars-b}
    A = -2\tau_2, \quad B = \ri(1+\tau_2)\sqrt{\frac{2\pi c}{3}},
\end{equation}
where $c$ is the central charge of the undeformed CFTs \footnote{For the Lee-Yang model, what controls the result is the ``effective central charge'' $c=2/5$ instead of the actual central charge $c=-22/5$. The two notions differ due to the non-unitary nature of the Lee-Yang model. In particular, the ``vacuum'' state created by the identity operator is not the lowest energy state. As a result, its contribution, which is related to the actual central charge, does not dominate the partition function in the high/low temperature limits, which is what defines the effective central charge. }:
\begin{equation}
\label{eq:pars-c}
    c=\begin{cases}
        1, &\text{free boson},\\  
        1/2, &\text{free fermion},\\
        2/5, &\text{Lee-Yang}.
    \end{cases}
\end{equation}
In addition, the parameter $\nu$ is found to be
\begin{equation}
\label{eq:pars-nu}
    \nu = \begin{cases}
        -1, &\text{free boson},\\
        -1/2, &\text{free fermion, Lee-Yang}.
    \end{cases}
\end{equation}
This is illustrated in Figs.~\ref{fig:s3bt1}, \ref{fig:s3bt4o5}, \ref{fig:s3bt9o8} for the $T\overline{T}$-deformed free boson, in Figs.~\ref{fig:s3ft1}, \ref{fig:s3ft3o4}, \ref{fig:s3ft4o5} for the $T\overline{T}$-deformed free fermion, and finally in Figs.~\ref{fig:s3LYt1}, \ref{fig:s3LYt4o5}, \ref{fig:s3LYt9o8} for the $T\overline{T}$-deformed Lee-Yang model.
In the process, we used the generalised Richardson transform, which, as explained in Appendix \ref{sec:Richardson} can remove higher $\mc{O}(n^{-k/2})$ corrections, to speed up the numerical convergence.

It is known from the resurgence theory that the large order behavior of perturbative coefficients are closely related to possible non-perturbative corrections, known as the resurgent relation.
As we explain in Appendix~\ref{sec:ResRel}, the asymptotic behavior of the type \eqref{eq:Zn-asymp} 
implies the non-perturbative corrections of the particular type
\begin{equation}\label{eq:np_predic}
    Z^{\text{np}} \sim \re^{-\frac{A}{\lambda}-\frac{B}{\sqrt{\lambda}}}\lambda^{-\nu}\sum_{k=0}^\infty b_k\lambda^{k/2}.
\end{equation}
We will further see in section~\ref{sec:Saddles} using the saddle point analysis of the integral form of the $T\overline{T}$-deformed partition function \eqref{eq:intRepZ} that $T\overline{T}$-deformed CFTs such as free boson, free fermion, and the Lee-Yang model indeed have non-perturbative corrections of this type, with the correct parameters given in \eqref{eq:pars-b}, \eqref{eq:pars-c}, and \eqref{eq:pars-nu}.

\subsection{Saddle-point analysis and resurgent properties}
\label{sec:Saddles}

In this section, we perform a detailed comparison between the resurgence properties of the exact expansion coefficients and the saddle-point contributions. To this end, we need to first compute the predictions for the resurgence results based on the saddle-point analysis. We begin with the integral representation of the $T\overline{T}$-deformed partition function: 
\begin{eqnarray}
\label{eq:integral_rep}
    \mathcal{Z}(\tau,\bar{\tau}|\lambda) = \frac{\tau_2}{\pi \lambda} \int_{\mathcal{H}_+} \frac{\rd^2 \zeta}{\zeta_2^2} K(\zeta,\bar{\zeta},\lambda)Z_{\text{CFT}}\left(\zeta,\bar{\zeta}\right),
\end{eqnarray}
with
\begin{equation}
    K(\zeta,\bar{\zeta},\lambda)= \exp\left(-\frac{|\zeta-\tau|^2}{\lambda \zeta_2}\right)\,. 
\end{equation}
For the purpose of explicit computations, we approximate the CFT partition function by the following simplified expression: 
\begin{align}
\label{eq:toy_model}
    Z_{\text{CFT}}(\zeta,\bar{\zeta}) \approx 
    &Z_{\text{TM}}(\zeta,\bar{\zeta})\\
    = 
    &\exp\left[-\frac{\pi c_L \ri}{6}\left(\zeta-\zeta^{-1}\right)+\frac{\pi c_R\ri}{6}\left(\bar{\zeta}-\bar{\zeta}^{-1}\right)\right]\nonumber
\end{align}
Let us explain why this approximation is relevant for our purpose. An important feature of the integral representation  (\ref{eq:integral_rep}) is that the expansion parameter $\lambda$ only appears in the $T\overline{T}$ kernel $K(\zeta,\bar{\zeta},\lambda)$. As a result, a naive saddle-point analysis would only involve this factor at the leading order. There is one caveat to this, that comes from the singularities of the CFT partition function $Z_{\text{CFT}}(\zeta,\bar{\zeta})$, near which the otherwise sub-leading free energy $F_{\text{CFT}} = -\ln{Z_{\text{CFT}}}$ diverges and can compete with the leading-order term from the $K(\zeta,\bar{\zeta},\lambda)$, the balancing of which can give rise to new saddle-points. Our approximation is proposed to capture the singular behavior of $Z_{\text{CFT}}$ near which the holomorphic and anti-holomorphic coordinates $(\zeta,\bar{\zeta})$ can approach either $0$ and $\infty$ independently. We comment that the chiral divergences of this nature, \emph{i.e.} as functions of $\zeta$ and $\bar{\zeta}$ independently, are features of the individual characters. They remain features of the partition function for rational CFTs which sums over only finitely many characters. Therefore the approximation is good for rational CFTs, which we focus on in this work. We leave the analysis for chaotic CFTs to future studies. Focusing on CFTs with $c_L=c_R = c/2$ and setting $\tau=\ri\tau_2$, the integral features an effective action of the form: 
\begin{equation}
\label{eq:EFFA}
    \mathcal{Z}(\tau,\bar{\tau}|\lambda) \sim  
    \frac{\tau_2 e^{\frac{2\tau_2}{\lambda}}}{\pi \lambda} \int_{\mathcal{H}_+} 
    \frac{\rd^2 \zeta}{\zeta_2^2} e^{-I_{\text{TM}}},
\end{equation}
with
\begin{equation}
    I_{\text{TM}}(\zeta_{1,2},\lambda) = \frac{1}{\lambda}\left(\frac{\zeta_1^2+\tau_2^2}{\zeta_2}+\zeta_2\right)-\frac{\pi c}{6}\left(\zeta_2+\frac{\zeta_2}{\zeta_1^2+\zeta_2^2}\right)\,.
\end{equation}

The saddle-points of this effective action can be solved exactly. As was alluded to before,  it consists of two classes of saddle-points: those produced solely by the kernel $K(\zeta,\bar{\zeta},\lambda)$ at the leading order in $\lambda$: 
\begin{equation}\label{eq:saddle_1}
    \zeta^*_1=0,\;\;\zeta^*_2 = \pm \sqrt{\frac{6\tau_2^2-\pi c \lambda}{6-\pi c \lambda}} \approx \pm \tau_2 +...
\end{equation}
and those produced by an interplay between the $T\overline{T}$ kernel $K(\zeta,\bar{\zeta},\lambda)$ and the singularities of (\ref{eq:toy_model}):
\begin{equation}\label{eq:saddle_2}
    \zeta^*_1 =  s''\sqrt{\frac{6\tau_2^2+ \pi s c\tau_2 \lambda}{\pi c\lambda}},\;\;\zeta^*_2 = \ri\tau_2 s'\sqrt{\frac{6}{\pi c\lambda}},
\end{equation}
with the choices
\begin{equation}
    s,s',s''=\pm 1\,.
\end{equation}
Regarding the second type (\ref{eq:saddle_2}), one can check that in the limit of $\lambda\to 0$, they behave like $\zeta^* \sim 1/\bar{\zeta}^* \to 0$ or $\bar{\zeta}^*\sim 1/\zeta^* \to 0$. This means that they are self-consistently located near which the approximation (\ref{eq:toy_model}) is valid, and therefore can be trusted to exist in real CFTs after the deformation, with appropriate corrections added. It is interesting to note that these saddle-points are analytically continued out of the Euclidean section $\zeta^\dagger = \bar{\zeta}$, and occur in the ``Lorentzian Regge'' regime.

The contributions from these saddle-points are summarized as the following. First of all, the contribution from the physical saddle with $\zeta^*_2\approx \tau_2$ in (\ref{eq:saddle_1}) contributes at the perturbative order $\lambda^0$:
\begin{align}
    \mathcal{Z}^{\text{pert}}(\tau,\bar{\tau}|\lambda) 
    &\sim  \exp\left(\frac{2\tau_2}{\lambda}-\frac{\sqrt{\left(6\tau_2^2-\pi c\lambda\right)\left(6-\pi c \lambda\right)}}{3\lambda}\right)\nonumber\\
    &\approx Z_{\text{CFT}}(\tau,\bar{\tau})+\mathcal{O}(\lambda)
\end{align}
We use the superscript ``pert'' to emphasize this is the perturbative part of the full partition function.
{Regarding the saddle-points contribution from those of our interest in (\ref{eq:saddle_2}), our goal is to carry out a thorough comparison up to higher orders in large $n$ with the resurgence property of the perturbative expansion coefficients $Z_n$ we analysed in section~\ref{sec:Large-order}.}
For this reason, we have computed the higher order corrections in small $\lambda$. They come from integrating over fluctuations about the saddle-points in expanding (\ref{eq:EFFA}) and including non-Gaussian interacting terms necessary for the corresponding order in $\lambda$. Their contributions are of the form: 
\begin{align}\label{eq:saddle_contributions_1}
    \mathcal{Z}^{\text{np}}(\tau,\bar{\tau}|\lambda) \sim &\lambda^{\frac{1}{2}(1+\alpha)}\exp\left(\frac{2\tau_2}{\lambda}+\ri\sqrt{\frac{2\pi c}{3\lambda}}K^{s,s'}\right)\nonumber\\
    &\left(1+ \frac{3\ri K^{s,s'}}{8\tau_2}\sqrt{\frac{3}{c}}\lambda^{1/2}+\mathcal{O}(\lambda)\right)
\end{align}
where the different saddles from the class (\ref{eq:saddle_2}) contribute with different 
\begin{equation}
    K^{s,s'}=s'\tau_2+ss'
\end{equation}
at the next-leading order. 
To be more precise we have also modified the contribution by putting in hand an additional exponent $\alpha$ that is theory dependent. It accounts for a possible additional power law factor of $\lambda^{\alpha/2}$ appended to (\ref{eq:toy_model}) in the Regge limit. For free boson we have $\alpha =1$, while for free fermion and Lee-Yang model we have $\alpha =0$. These exponents can be extracted from the transformation properties of the $\eta(\tau)$ under the inversions $\tau\to -1/\tau$: 
\begin{align}
    \eta(-1/\tau) \sim \sqrt{-i\tau} \eta(\tau)
\end{align}
Due to this, near the chiral singularities the toy-model for the free-boson could be improved by taking the pre-factor of this transformation into account: 
\bea
Z^{\text{B}}_0 &=& \frac{\sqrt{-\ri\bar{\zeta}}}{\sqrt{\zeta_2}\eta(\zeta) \eta(-1/\bar{\zeta})}\nonumber\\
&\approx & \left(-\ri\bar{\zeta}/\zeta_2\right)^{1/2}\exp\left[-\frac{\pi c \ri}{12}\left(\zeta+\bar{\zeta}^{-1}\right)\right] 
\eea
while that of other minimal models including the free-fermion and Lee-Yang model are unaffected, because their characters $\chi_{r,s}$ transform only by a constant matrix under inversion. Plugging the saddle-point behavior $\zeta_2 \sim 1/\bar{\zeta} \sim \lambda^{-1/2}$ then gives the exponent $\alpha$. We comment that $\alpha$ is the first model-specific correction to the toy model, and it affects the asymptotic behavior of the $s_n$ at the order $1/n$. Higher order terms in $s_n$ encode more model-dependent data, which is beyond the validity of the effective model. We will come back to this point at the end of this section. Notice that the expansions in the non-perturbative sectors are controlled by powers of $\lambda^{1/2}$, which slightly differ from the standard Gevrey-1 type usually encountered in QFTs.   

Next we connect the general form of the non-perturbative contribution to the partition function that we have encountered to the resurgence properties of the perturbative expansion coefficients $Z_n$ at large orders. Suppose the partition function has a nonperturbative component of the form
\begin{equation}
\label{eq:Znp}
    Z^{\text{np}}(\lambda) = \lambda^{-\nu}\re^{-A/\lambda-B/\sqrt{\lambda}}\sum_{k=0}^\infty b_k \lambda^{k/2},
\end{equation}
in accord with \eqref{eq:saddle_contributions_1}, by a standard resurgence analysis, as explained in Appendix~\ref{sec:ResRel}, it will contribute to the following asymptitc behavior of $Z_n$
\begin{align}
\label{eq:b1A}
  &Z_n \sim 
  \re^{\frac{B^2}{8A}}b_0\sqrt{\frac{\pi}{3}}
  \frac{\Gamma(n+\nu)}{A^{n+\nu}}
  \re^{B\sqrt{\frac{n+\nu}{A}}}\\\nonumber
  &\left(1 +\left(\frac{b_1A^{1/2}}{b_0}+
      \frac{12AB-B^3}{96A^{3/2}}
    \right)\frac{1}{\sqrt{n+\nu-1/2}}+\mc{O}(1/n)\right).  
\end{align}
In our case at hand, by comparing \eqref{eq:Znp} and \eqref{eq:saddle_contributions_1}, we find
\begin{align}
\label{eq:asymp_param}
    A = -2\tau_2,\;\;&B= \ri\sqrt{\frac{2\pi c}{3}} K^{s,s'},\;\;\nu = -\frac{1}{2}(1+\alpha),\;\;\;\\\nonumber
    &
    K^{s,s'}=s'\tau_2+ss'.
\end{align}
Notice that we have $\nu = -1$ for the free boson and $\nu = -1/2$ for the free fermion and the Lee-Yang model. The saddles from (\ref{eq:saddle_2}) labelled by $(s,s')$ make contributions to $Z_n$ that are equal at the leading order in large $n$. However, their contributions at the next-leading order still differ by the exponentially in $\sqrt{n}$ factor: 
\begin{equation}
    Z^{s,s'}_n\propto \re^{(s'\tau_2+ss')\sqrt{2\pi cn/3\tau_2}}
\end{equation}
As a result, what actually controls the asymptotic growth of $Z_n$ is the saddle with the largest positive exponent, \emph{i.e.} that with
\begin{equation}
    s=s'=1,
\end{equation}
leading to the large order behavior \eqref{eq:Zn-asymp} with the parameters \eqref{eq:pars-b}, \eqref{eq:pars-c}, \eqref{eq:pars-nu}.
This is confirmed by numerical studies of $nZ_n/Z_{n+1}$ in section~\ref{sec:Large-order}.



We end this section by making a few comments regarding the comparison. 
\begin{itemize}
\item The coefficients for the first three orders in the large $n$ expansion of $nZ_n/Z_{n+1}$ match very well between the exact results and the saddle-point prediction (\ref{eq:resurgence_prediction}). This is the case for the free boson, the free fermion, as well as the Lee-Yang model. On the other hand, the exact results exhibit interesting transient oscillatory features before settling down to the predicted values. It is reasonable to suppose that they reflect the full modular properties of the seed partition functions, which are much richer than the toy model (\ref{eq:toy_model}). In particular, there are additional saddle-points of (\ref{eq:integral_rep}) that are analogous to (\ref{eq:saddle_2}): they arise by balancing the $T\overline{T}$ kernel with other singularities of the seed CFT partition function -- the images under modular transformations of the singularities captured by the toy model. In Appendix~\ref{sec:modular_images} we perform a qualitative analysis of their effects. Their contributions to the expansion coefficients are found to be sub-leading at large order $n$, and oscillates with $n$ -- compatible with the qualitative features of the exact results. It is interesting to analyze these in greater details in future investigations.  

\item It is interesting to observe that it takes larger order $n$ for the exact results of the free fermion and the Lee-Yang model to stop oscillating and settle down to the saddle-point prediction,  compared to that of the free boson. We suspect that it is related to the ways in which the exact seed partition functions differ non-perturbatively from that of the toy model near the saddle-points. For example one can estimate such differences near the saddle-point:
\begin{equation} \label{eq:regge_saddle_eg}
\zeta \approx \sqrt{\frac{\pi c}{24\lambda}}\tau_2,\;\;\bar{\zeta} \approx \sqrt{\frac{\pi c \lambda}{24}} 
\end{equation}
They are controlled by: 
\bea
    &&Z^{\text{B}}_0 \left(\zeta,\bar{\zeta}\right)= \frac{1}{\sqrt{\tau_2}\eta(\zeta)\bar{\eta}(\bar{\zeta})} \\\nonumber
    &&\sim Z_{\text{TM}}\left(\zeta, \bar{\zeta}\right)\Big(1 + \mathcal{O}\left(q_{\text{B}}\right)+\mathcal{O}\left(\bar{q}_{\text{B}}\right)\Big),\\\nonumber
    &&q_{\text{B}}=e^{2\pi \ri \zeta_B},\;\;\bar{q}_{\text{B}} = e^{-2\pi \ri/\bar{\zeta}_{\text{B}}},
\eea
for the free boson and
\bea
    &&Z^{\text{F}}_0 \left(\zeta,\bar{\zeta}\right)= \sum^4_{i=2}\sqrt{\frac{\theta_i(\zeta)\bar{\theta}_i(\bar{\zeta})}{\eta(\zeta)\bar{\eta}(\bar{\zeta})}} \\\nonumber
    &&\sim Z_{\text{TM}}\left(\zeta, \bar{\zeta}\right)\Big(1 + \mathcal{O}\left(q^{1/2}_{\text{F}}\right)+\mathcal{O}\left(\bar{q}^{1/2}_{\text{F}}\right)\Big),\\\nonumber
    &&q_{\text{F}}=e^{2\pi \ri \zeta_F},\;\;\bar{q}_{\text{F}} = e^{-2\pi \ri/\bar{\zeta}_{\text{F}}},
\eea
for the free fermion, where $\zeta_{\text{B,F}}$ and $\bar{\zeta}_{\text{B,F}}$ are  given by (\ref{eq:regge_saddle_eg}) upon setting $c=1$ and $c=1/2$ respectively. We can roughly estimate their effects in the large order coefficients $Z_n$ by plugging the saddle point (\ref{eq:large_order_saddle}) for $\lambda_+ = x_+^2$ at large $n$. It is then easy to discover that: 
\begin{align}
    &q_{\text{B}} \sim \re^{-\tau_2\sqrt{\frac{\pi^3 n}{12\tau_2}}}  \ll \re^{-\tau_2\sqrt{\frac{\pi^3 n}{96\tau_2}}} = q_{\text{F}}^{1/2},\;\; \\
    &\bar{q}_{\text{B}} \sim \re^{-\sqrt{\frac{\pi^3 n}{12\tau_2}}}  \ll \re^{-\sqrt{\frac{\pi^3 n}{96\tau_2}}} = \bar{q}_{\text{F}}^{1/2}
\end{align}
In other words, the effects from the deviation between the exact results and the toy model decay much faster in the free boson than that in the free fermion. Similar comparison can also be made between the free boson and the Lee-Yang model. This may provide an explanation for the slower convergence of the exact results to the predicted values in the free fermion and the Lee-Yang model. 

\item In principle, the comparison can be  extended further to the higher order coefficients of $s_n$ in the large $n$ expansion. For example, one can compare the $n^{-3/2}$ term in $s_n$, whose predicted coefficient involves $b_0/b_1$ in (\ref{eq:asymp_param}). However, at this order we found that the exact results do not match the saddle-point prediction. To explain this mismatch, one can show that near (\ref{eq:saddle_2}) the approximation (\ref{eq:toy_model}) is accurate up to order $\mathcal{O}(\lambda^{1/2})$ compared to the exact CFT partition function. Corrections of this order need to be considered in order to push the match further. One can further show that such corrections begin to affect $s_n$ precisely at the order $n^{-3/2}$. The mismatch therefore has a well-understood origin and fix: we simply need to improve the approximation (\ref{eq:toy_model}) to higher orders -- a task we shall leave for future investigations. 

\item One might wonder whether the other saddle-point in (\ref{eq:saddle_1}): $\zeta^*_1 = 0,\; \zeta^*_2 \approx - \tau_2$ is excluded from the resurgence analysis. In principle, this saddle-point does affect the asymptotic behavior of the perturbative coefficients $Z_n$ just like other saddle-points. In practice, one can estimate its contribution from the leading order action, which scales like $\sim 4\tau_2/\lambda$. Via the standard resurgence analysis, the contribution to the asymptotic coefficients $Z_n$ scales like $\sim n! (-4\tau_2)^{-n}$ at large $n$. This effect is exponentially small compared to the effects from other saddle-points, and so is eclipsed in the numerical results. 

\end{itemize}

In conclusion, the comparison we have performed provides strong evidence that the resurgence properties we observe in the exact expansion coefficients are indeed controlled by the saddle-points of the nature (\ref{eq:saddle_2}).

\subsection{Stokes' phenomenon and negative $\lambda$}
\label{sec:Stokes}
In this section, we study the Stokes' phenomenon associated with the saddle-points we have identified from the resurgence analysis. The readers can refer to Appendix~\ref{sec:StokesReview} for a brief review of the Stokes' phenomenon. Our goal for studying the Stokes' phenomenon is as follows. In general, the non-perturbtive contribution from a particular saddle-point does not necessarily contribute for all complex values of the coupling constant $\lambda$. In which region of the complex $\lambda$-plane is it present is determined by studying the Stokes' phenomenon. This is important for analyzing the $T\overline{T}$-deformed theory with both $\lambda>0$ and $\lambda<0$ -- they are related by a rotation of $\lambda$ in the complex plane. 

Based on the effectiveness in capturing the non-perturbative effects, we shall use the approximation (\ref{eq:toy_model}) for analyzing the Stokes' phenomena.  In particular, we shall apply the analysis to the integral representation (\ref{eq:integral_rep}), with $Z_{\text{CFT}}$ approximated by the toy model (\ref{eq:toy_model}). Along the positive real axis $\lambda = |\lambda|>0$ and for temperatures satisfying $\tau_2 > \tau^c_2 = \sqrt{\pi c\lambda/6}$, it can be checked that the integration contour $\mathcal{H}_+$ decomposes into only the Lefschetz thimble $\mathcal{J}_p$ through the physical saddle-point  $\zeta^*_p = (\zeta^*_1,\zeta^*_2)$ in the class (\ref{eq:saddle_1}): 
\be 
\zeta^*_1 = 0,\;\;\zeta^*_2 = \sqrt{\frac{6\tau^2_2-\pi c \lambda}{6-\pi c \lambda}} \approx \tau_2+...
\ee
In fact for $\lambda>0$, the integration contour $\mathcal{H}_+$ actually coincides with $\mathcal{J}_p$, giving: 
\bea\label{eq:PF_positive}
\mathcal{H}_+ &=& \mathcal{J}_p\rightarrow\mathcal{Z}(\tau,\bar{\tau}|\lambda) = \mathcal{Z}_p(\tau,\bar{\tau}|\lambda) \sim e^{-I_{\text{TM}}(\zeta^*_p)}
\eea
with
\begin{align}
I_{\text{TM}}(\zeta^*_p)&=& \frac{1}{3\lambda}\sqrt{\left(6\tau_2^2-\pi c \lambda\right)\left(6-\pi c \lambda\right)}-\frac{2\tau_2}{\lambda}
\end{align}

\begin{figure}
    \centering
   {\includegraphics[width=0.45\linewidth]
   {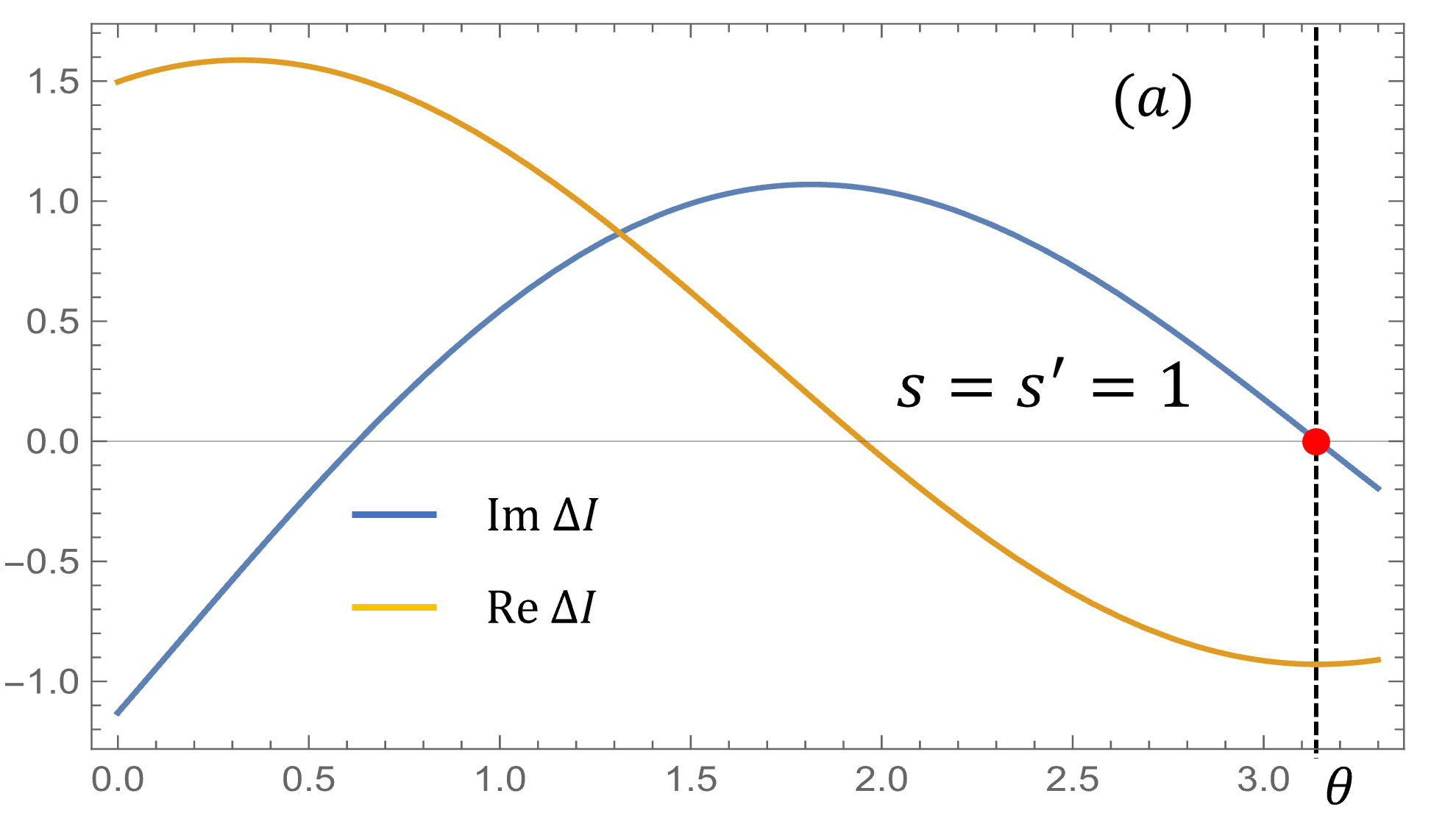}}
   {\includegraphics[width=0.45\linewidth]{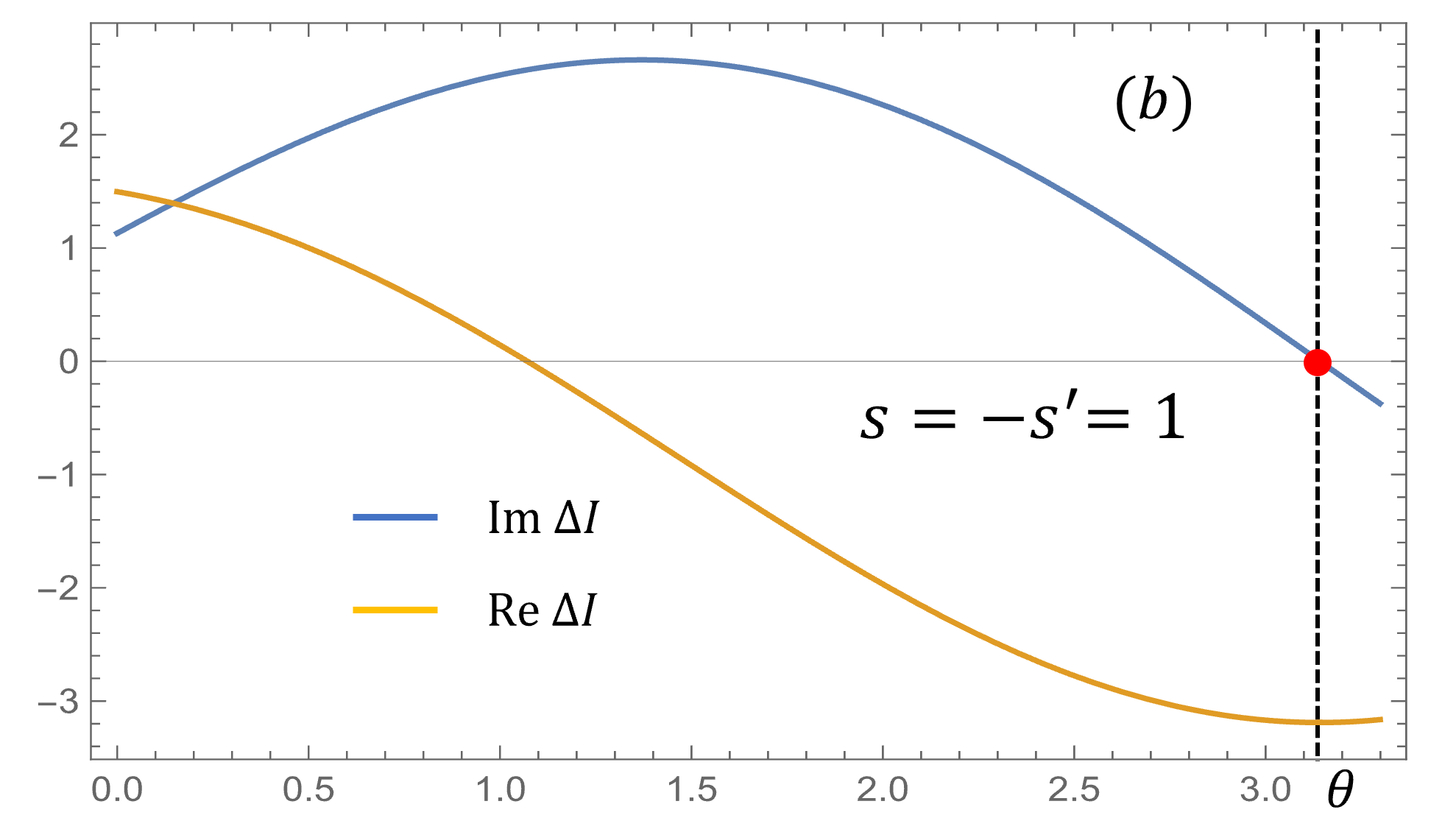}}
   {\includegraphics[width=0.45\linewidth]{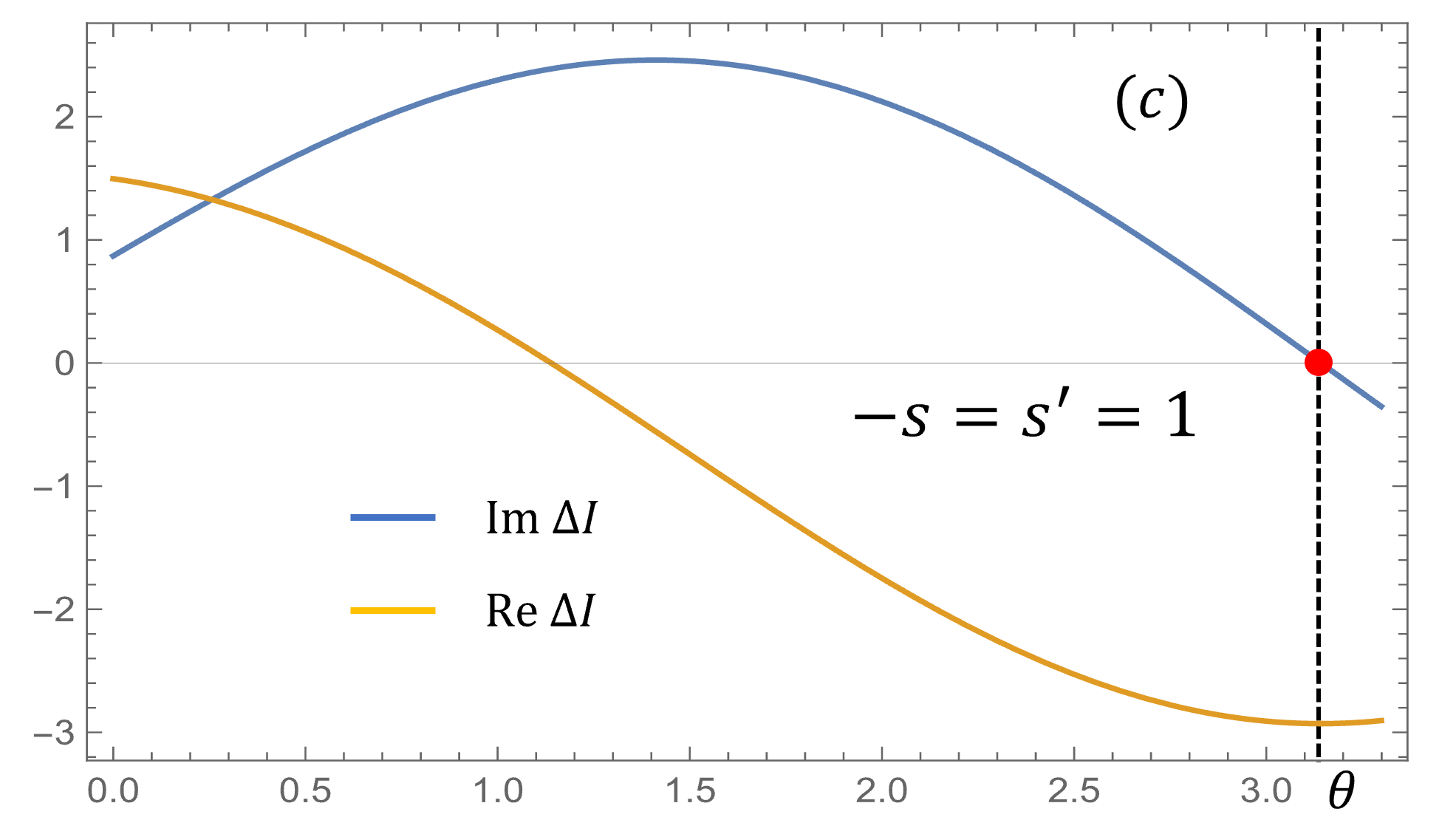}}
   {\includegraphics[width=0.45\linewidth]{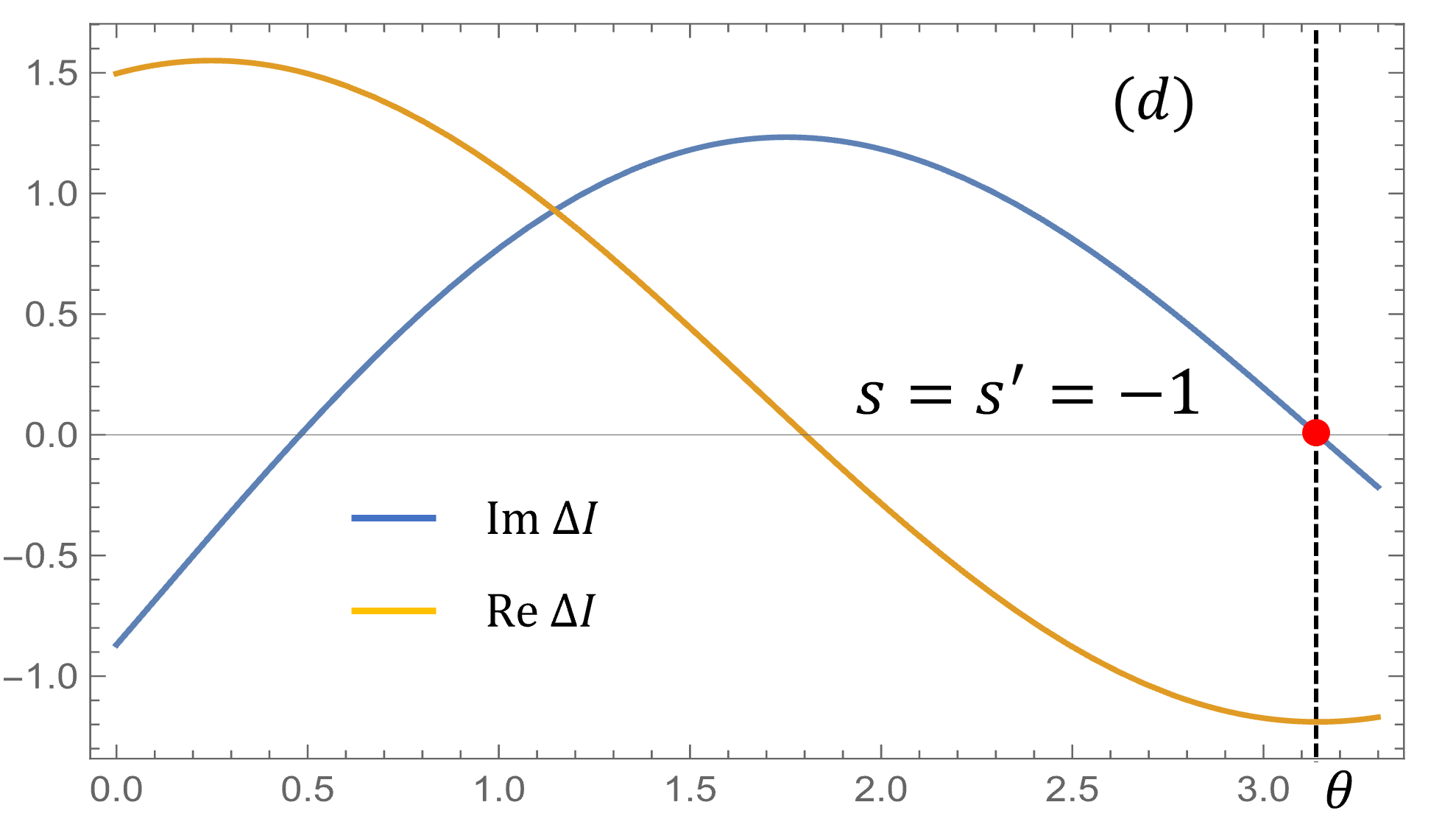}}
    {\includegraphics[width=0.45\linewidth]{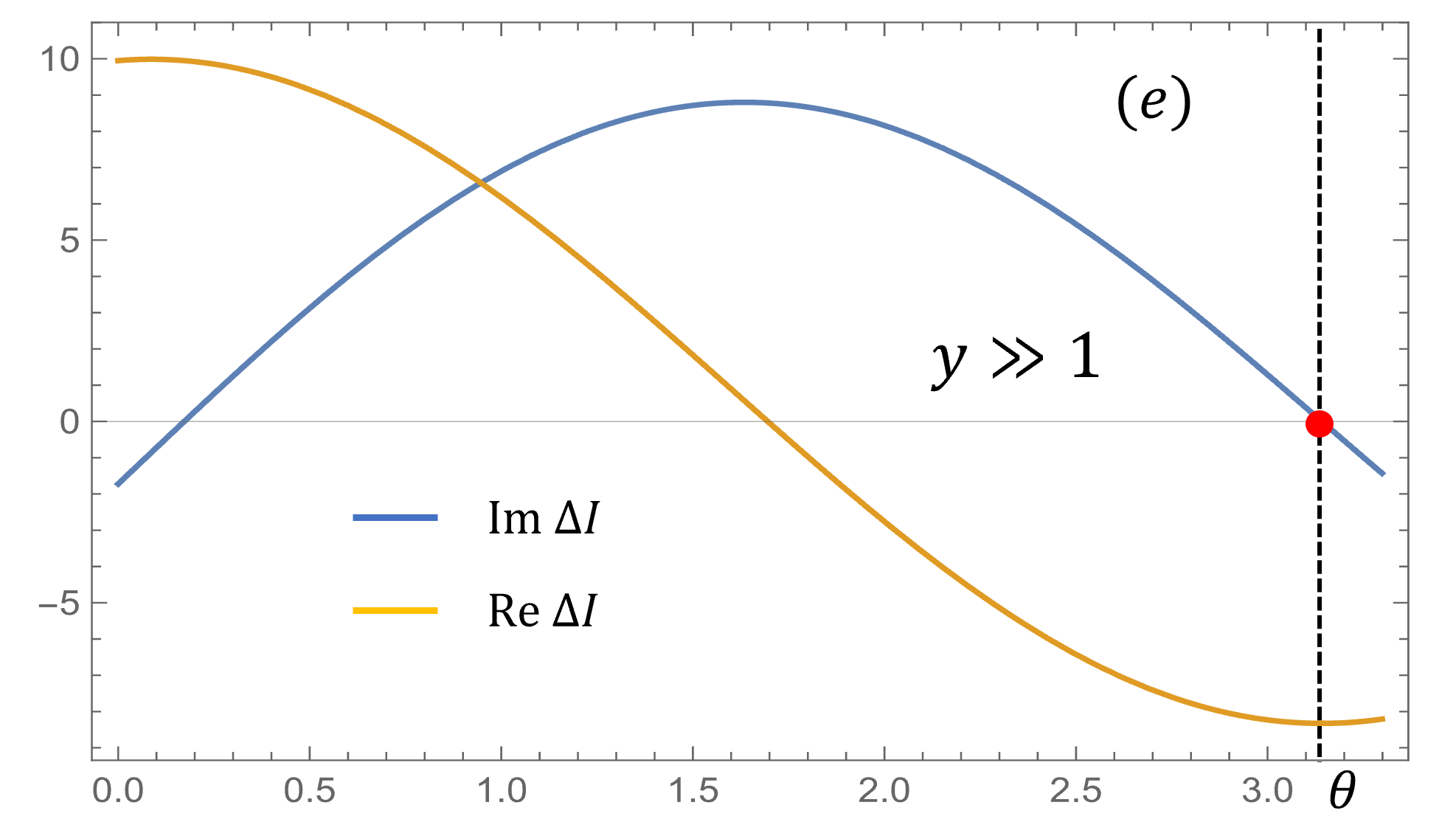}}
    \caption{Plots of $\text{Im} \Delta I$ (blue) and $\text{Re} \Delta I$ (yellow) featuring a Stokes phenomenon at $\theta=\pi$ (red dot) for the saddle-points $\zeta^*_{s,s'}$. The re-scaled temperature is of order 1 for $y=1.8$ and $|\lambda| =10^{-2}$ for (a)-(d); at low temperatures $y\gg 1$, the plots for distinct $s,s'$ become almost identical, shown in (e) for $y=10$ and $\lambda = 10^{-2}$. }
    \label{fig:Stokes_pheno}
\end{figure}

In particular, the contributions from all the other saddle-points are absent for positive $\lambda$. Notice that had they been present, they would amount to exponentially larger contributions of the order $e^{2\tau_2/|\lambda|}$ or $e^{4\tau_2/|\lambda|}$. As $\lambda = e^{i\theta}|\lambda|$ rotates in the complex $\lambda$-plane while kept at small modulus $|\lambda|<1$, the Lefschetz thimbles deform accordingly. Our focus is the Stokes' phenomenon associated with the saddle-points from (\ref{eq:saddle_2}). Based on the general discussion in Appendix~\ref{sec:StokesReview}, this simply requires us to identify the occurrences of the condition (\ref{eq:Stokes_pheno}) on the complex $\lambda$-plane. Let us label by $\zeta^*_{s,s',s''}$ on of the saddle-points from (\ref{eq:saddle_2}): 
\bea 
\zeta^*_{s,s',s''} &=& \left(\zeta^*_1,\zeta^*_2\right),\;\;\zeta^*_1 =  s''\sqrt{\frac{6\tau_2^2+ \pi s c\tau_2 \lambda}{\pi c\lambda}}\nonumber\\
\zeta^*_2 &=& \ri\tau_2 s'\sqrt{\frac{6}{\pi c\lambda}},\;\;s,s',s''= \pm 1
\eea
The saddle-contribution from $\zeta^*_{s,s',s''}$ is given by: 
\bea
\mathcal{Z}_{s,s',s''}(\tau,\bar{\tau}|\lambda)&\sim & e^{-I_{\text{TM}}\left(\zeta^*_{s,s',s''}\right)}\nonumber\\
I_{\text{TM}} \left(\zeta^*_{s,s',s''}\right) &=& -\ri\sqrt{\frac{2\pi c}{3\lambda}}B - \frac{2\tau_2}{\lambda},\;\;B = s'\tau_2+ss'\nonumber
\eea
Since it does not depend on $s''$, we omit the index $s''$ from now on. The condition (\ref{eq:Stokes_pheno}) for the Stokes' phenomenon associated with the saddle-point $\zeta^*_{s,s'}$ amounts to that: 
\be
\text{Im}\;\Delta I = 0,\;\;\;\text{Re}\;\Delta I<0
\ee
The action difference $\Delta I$ at the leading orders in small $\lambda$ can be written as: \footnote{In performing the following Stokes analysis, we work in the semi-classical limit of the high temperature regime: $\tau_2\sim \tau^c_2 \sim\sqrt{\lambda}$, in which we define $y = \tau_2/\tau^c_2$ as the effective parameter; the low temperature regime is then represented by $y\gg 1$. In doing so, we made the approximation $(6\tau^2_2-\pi c \lambda)(6-\pi c \lambda) \approx 36\tau^2_2- 6\pi c \lambda$ for the saddle-point action of $\zeta^*_p$. This is valid to the leading order in both the high and low temperature regimes.} 
\bea
\Delta I &=& I_{\text{TM}}\left(\zeta^*_p\right) - I_{\text{TM}}\left(\zeta^*_{s,s'}\right)\nonumber\\
&=& 2\lambda^{-1}\sqrt{\tau_2^2-\frac{\pi c\lambda}{6}}+\ri \sqrt{\frac{2\pi c }{3\lambda}} B +...\nonumber\\
&=& \frac{2\tau^c_2}{|\lambda|} \left(e^{-\ri\theta}\sqrt{y^2-e^{\ri\theta}}+\ri B e^{-\ri\theta/2}\right)+...
\eea
for the complex $\lambda = e^{\ri\theta}|\lambda|$ and the re-scaled temperature $y$ defined by $\tau_2 =  y\;\tau^c_2$. The condition (\ref{eq:Stokes_pheno}) for the Stokes' phenomenon involving $\zeta^*_{s,s'}$ can therefore be written explicitly in terms of phase $\theta$: 
\bea\label{eq:Stokes_pheno_2} 
&&\sqrt{A} \sin{\left(\frac{\Theta}{2}+\theta\right)}-B \cos{\left(\frac{\theta}{2}\right)}=0\nonumber\\
&&\sqrt{A}\cos{\left(\frac{\Theta}{2}+\theta\right)}+B \sin{\left(\frac{\theta}{2}\right)}<0\nonumber\\
&&A=\sqrt{y^4-2y^2 \cos{\theta}+1},\;\;\tan{\Theta} = \frac{\sin{\theta}}{y^2-\cos{\theta}}
\eea
It can be checked that for $y>1$ and starting from $\theta =0$, the Stokes' phenomena occurs at $\theta=\pi$ for any choice of $s,s'$. For an illustration, in Fig.~\ref{fig:Stokes_pheno} we plot $\text{Im}\; \Delta I$ and $\text{Re}\; \Delta I$ as functions of $\theta$, both for low temperatures $y\gg 1$ or near the critical temperature $y\sim \mathcal{O}(1)$.

We conclude as $\lambda$ is analytically continued from $\lambda>0$, it receives non-perturbative contributions from the saddle-points in (\ref{eq:saddle_2}) when $\lambda\to -|\lambda|<0$. We can therefore write the continued partition function in terms of the positive parameter $|\lambda|$ as: 
\bea\label{eq:PF_negative}
&&\mathcal{Z}(\tau,\bar{\tau}|-|\lambda|) = \mathcal{Z}_p(\tau,\bar{\tau}|-|\lambda|) + \sum_{s,s'} \mathcal{Z}_{s,s'}(\tau,\bar{\tau}|-|\lambda|) \nonumber\\
&\sim & e^{-\frac{2\tau_2}{|\lambda|}+\frac{1}{3|\lambda|}\sqrt{\left(6\tau_2^2+\pi c |\lambda|\right)\left(6+\pi c |\lambda|\right)}} + e^{-\frac{2\tau_2}{|\lambda|}+\sqrt{\frac{2\pi c}{3|\lambda|}}} 
\eea
To clarify, in writing the non-perturbative corrections $\mathcal{Z}_{s,s'}$ in the second line we have only kept the contributions from the saddle-points with $ss'=1$ because they dominate over the others at the order of $\mathcal{O}(|\lambda|^{-1/2})$ in the exponent. In addition, we only kept the leading order term in the coefficient of $\tau_2$ in the exponent of the non-perturbative correction. Notice that while (\ref{eq:PF_positive}) as a function of the inverse temperature $\tau_2$ has a physical singularity at $\tau^c_2>0$, the expression (\ref{eq:PF_negative}) is now singularity-free for real and positive $\tau_2$. So we can use this expression to obtain an analytic continuation of $\mathcal{Z}(\tau,\bar{\tau},-|\lambda|)$ above the critical temperature $T_c=1/\tau^c_2$, making it well-defined for all physical $\tau_2>0$. 

Interestingly, we see that the $T\overline{T}$-deformed partition function for $\lambda<0$ is not simply the analytic continuation of the $\lambda>0$ counter-part. The Stokes' phenomenon has given rise to non-perturbative effects at $\lambda<0$ that are absent for $\lambda>0$. Such explicit asymmetry between the positive and negative $\lambda$ theories resonates well with the expectation that they exhibit drastically distinct physics, especially in their spectral properties. While the $\lambda>0$ theory features a stringy Hagedorn growth in the spectral density, the $\lambda<0$ theory encounters the issue of complex spectrum in the UV, respecting unitarity then calls for a UV cut-off in the energy spectrum. In particular, the spectral properties for positive and negative $\lambda$ theories are manifestly not related by a simple analytic continuation.

It is therefore tempting the explore the implications of (\ref{eq:PF_negative}) regarding the spectral property of the $\lambda<0$ theories, and see whether they align with physical expectations. The two terms in (\ref{eq:PF_negative}) point to distinct physical properties of the theory at $\lambda<0$. The first term is the naive analytic continuation of the partition function at $\lambda>0$. This term alone points to complex energies similarly obtained by analytically continuing the spectrum of $\lambda>0$. The second term alone, coming from a distinct class of saddle-points, is instead compatible with imposing a UV cut-off at $\Lambda=2|\lambda|^{-1}$: 
\begin{eqnarray} 
Z(\tau,\bar{\tau},-|\lambda|) &\sim & \int^{\Lambda} dE\; e^{-\tau_2 E+ S(E)}\nonumber\\
&\sim &  e^{-\frac{2\tau_2}{|\lambda|}+S(2/|\lambda|)} +...
\end{eqnarray}
In the second line we performed an integration by parts and collected the boundary term.\footnote{The bulk term can be shown to produce only sub-leading corrections to the boundary term, and can be made to agree with the non-perturbative term in (\ref{eq:PF_negative}) by including corrections to the expressions of the cut-off $\Lambda$ and entropy $S(E)$. } It matches the second term for a Cardy-like entropy: $S(E)=\sqrt{\pi E c/3}$.   

In other words, our analysis show that the two distinct saddle-point contributions to the partition function, one of which enters via the Stokes' phenomenon, represent two distinct options to treat the $\lambda<0$ theory according to the spectral flow equation: (i) allowing complex spectrum and losing unitarity; (ii) keeping unitarity by imposing  a UV cut-off. However, we should emphasize that our analysis has not settled the puzzles concerning the $\lambda<0$ theory decisively. The very fact that two terms representing complementary pictures are simultaneously present in (\ref{eq:PF_negative}) means that we have not reconciled them into a coherent resolution. This is beyond the scope of our paper. Nonetheless, we end this section by making some comments regarding the challenges associated with a full resolution.  

One of the main challenges is the singularity at $\tau_2 = \tau^c_2$ in the positive $\lambda$ partition function (\ref{eq:PF_positive}), which is reminiscent of the Hagedorn singularity in string theory \cite{PhysRevLett.25.895,Fubini:1969qb,Hagedorn:1965st}. Its impact on the Stokes' phenomenon is unclear. Although by first restricting to $\tau_c>\tau^c_2$ and then analytically continuing to negative $\lambda$, we arrived at an expression that is non-singular for all $\tau_2$, a more rigorous treatment should proceed by first finding the correct partition function for all $\tau_2$ at positive $\lambda$, and then analytically continuing $\lambda$ to the negative real-axis for the full expression. Doing this inevitably requires resolving the Hagedorn singularity, which is by itself an extremely interesting yet challenging endeavor. Motivated by the understanding in string theory \cite{Sathiapalan:1986db,Kogan:1987jd,OBrien:1987kzw,ATICK1988291}, it possibly involves studying all the winding mode contributions (see e.g. \cite{Benjamin2023} for the role it could play in $T\overline{T}$-deformed CFTs), and even contributions from the target space of non-torus topology, in the flat JT-gravity or non-critical string theory formulation of the $T\overline{T}$-deformation \cite{Dubovsky:2018bmo,Callebaut:2019omt,Hashimoto:2019wct}, from which the integral representation (\ref{eq:integral_rep}) is derived. We leave this for the future investigations. 

Another major challenge is to identify and incorporate other non-perturbative effects. Apart from the ones considered in this section, there is also the non-perturbative contribution from the remaining saddle of (\ref{eq:saddle_1}), i.e. that of $\zeta^*_2 \sim -\tau_2$. We did not consider it because its effect on the asymptotic behavior of the perturbative coefficients is too suppressed to be observable at large orders, yet they can still make a contribution to the partition function. It is easy to check that the associated Stokes' phenomenon proceeds similarly: its contribution is absent at $\lambda>0$, and enters at $\lambda<0$, where it is suppressed by $e^{-4\tau_2/|\lambda|}$ and so is negligible to both terms in (\ref{eq:PF_negative}). As was alluded to before and discussed in  Appendix~\ref{sec:modular_images}, for actual CFTs there are more saddle-points of similar nature as the ones we considered. They occur close to other singularities of the seed CFT partition function, whose existences are dictated by the full modular invariance. There could be infinitely many such saddle-points, a typical contribution of which takes the form (\ref{eq:more_saddle}) and is in general complex. To fully understand the theory at $\lambda<0$ requires studying their Stokes' phenomenon and possibly re-summing their effects as $\lambda$ rotates in the complex-plane from the positive real axis to the negative real axis. We leave this also to the future studies.

\section{Comments on holography}
\label{sec:holography}
In this section, we comment on the possible implication of our result on holography. We discuss both the double-trace and single-trace $T\overline{T}$-deformations.

\subsection{Double-trace $T\overline{T}$ deformation} 
In a typical double-trace $T\overline{T}$-deformed AdS$_3$/CFT$_2$ set-up, the boundary theory is a $T\overline{T}$-deformed holographic CFT;  the bulk theory is Einstein gravity in AdS$_3$ with modified mixed boundary condition, which can be interpreted as a ``cut-off'' geometry (for $\lambda<0$) or ``glue-on'' geometry (for $\lambda>0$) in the pure gravity sector. Our analysis in the previous section shows that as we analytically continue $\lambda$ in the complex plane from $\lambda>0$ to $\lambda<0$, new non-perturbative contributions emerges due to Stokes' phenomenon.  In \cite{Aharony:2018bad} the possible non-perturbative ambiguity for $\lambda<0$ has already been noticed. In the current work, we have indeed found such a non-perturbative contribution from a concrete resurgence analysis. In principle a thorough analysis for all the Stokes poles on the complex $\lambda$-plane will uncover all the non-perturbative contributions. If our analysis can be appropriately extended to double-trace $T\bar{T}$-deformed holography, it suggests that for the $\lambda<0$ theories there will be new non-perturbative heavy objects that contribute to the gravity partition function. It would be interesting to find these contributions from the AdS side.

However, we should mention that strictly speaking our saddle-point analysis only applies to rational CFTs. The singularity structure of their partition functions is clear, making it easy to identify and study the saddle-points explicitly. On the other hand, holographic CFTs dual to local Einstein gravity in AdS$_3$ are generally expected to be chaotic with large central charge $c$. The singularities structure of their partition functions is much more opaque. Correspondingly, it is difficult to perform the same saddle-point analysis as is done in this paper. Despite the technical challenge, we believe that the general philosophy of our analysis may still help for studying non-perturbative effects in $T\overline{T}$-deformed holography. In particular, one can consider the potential interplay between the $T\overline{T}$ kernel and emergent features of the holographic CFT partition function at large $c$, the latter of which can be supplied from the gravity side. We leave these for future investigations. 

\subsection{Single-trace $T\overline{T}$ deformation} 
For the single-trace $T\overline{T}$-deformation, one can take a top-down approach. For concreteness, we focus on the holographic duality pair between the $k=1$ tensionless string on AdS$_3\times$S$^3\times\mathbb{T}^4$ and the symmetric orbifold CFT Sym$^N(\mathbb{T}^4)$. The single-trace $T\overline{T}$-deformation of both sides have been proposed in \cite{Dei:2024sct}, based on previous works \cite{Giveon:2017myj,Giveon:2017nie,Eberhardt:2018ouy,Giveon:1998ns}. In the bulk, the holographic dual corresponds to a marginal double current deformation on the worldsheet. The authors of \cite{Dei:2024sct} checked the deformed partition function and find a perfect match on both sides. They proposed an interesting non-perturbative completion of the deformed partition function from the bulk. The key proposal is taking into the so-called ``negative winding modes''. A similar proposal was put forward from a different perspective \cite{Benjamin:2023nts}. It is interesting to compare this proposal to our result and see if it emerges naturally from resurgence.\par

A full-fledged resurgence analysis requires considering symmetric product orbifolds, which is beyond the scope of the current work. However, note that the winding sector $w=1$ coincides with the $T\overline{T}$-deformed single copy CFT and with the current result we can already make meaningful comparisons with the first negative winding sector $w=-1$. To this end, we start with the spectrum. The $T\overline{T}$-deformed energy is obtained by solving Burgers' equation. For the CFT, the Burgers' equation can be recast as the following quadratic equation
\begin{align}
2\mathcal{E}_n(\lambda)+\lambda\pi R\left(\mathcal{E}_n(\lambda)^2-P_n^2\right)=2E_n
\end{align}
There are two solutions to this equation
\begin{align*}
\mathcal{E}_n^{\pm}(\lambda)=\frac{1}{\pi\lambda R}\left(-1\pm\sqrt{1+2\lambda\pi R E_n+\lambda^2\pi^2R^2P_n^2}\right)
\end{align*}
We usually take $\mathcal{E}_n^+(\lambda)$ as the deformed energy because it has a well-defined limit in $\lambda\to0$.
\paragraph{The negative branch} The negative branch $\mathcal{E}_n^-(\lambda)$ have several problems. In the limit $\lambda\to 0_+$, $\mathcal{E}_n^-(\lambda)$ takes large negative values for large $E_n$ and/or $P_n$, including such states in the spectrum will lead to instabilities. Alternatively, their contribution to the partition function
\begin{align}
\mathcal{Z}^-(\tau,\bar{\tau}|\lambda)\sim\sum_n e^{-2\pi\tau_2\mathcal{E}_n^-(\lambda)}
\end{align}
is divergent for $\lambda\to 0_+$. 

\paragraph{Negative winding} However, if we take a negative sign out of $\mathcal{E}_n^-(\lambda)$ and consider the energy
\begin{align}
\mathcal{E}_n^{\text{n.p.}}(\lambda)=&\,-\mathcal{E}_n^-(\lambda)\\\nonumber
=&\,\frac{1}{\pi\lambda R}\left(1+\sqrt{1+2\lambda\pi RE_n+\lambda^2\pi^2R^2 P_n^2}\right)
\end{align}
Now the energy $\mathcal{E}_n^{\text{np}}(\lambda)$ is positive, and its contribution to the partition function
\begin{align}
\mathcal{Z}^{\text{np}}(\tau,\bar{\tau}|\lambda)\sim \sum_n e^{-2\pi\tau_2 R\mathcal{E}_n^{\text{np}}(\lambda)}
\end{align}
vanish in the limit $\lambda\to0_+$ and can be viewed as a non-perturbative contribution. The states $(\mathcal{E}_n^{\text{np}}(\lambda),-P_n)$ are the negative winding modes $w=-1$ \cite{Dei:2024sct}. The proposal is that we take into account such states as well as analogous states for $w=-2,-3,\ldots$ in the deformed partition function.\par

Such a non-perturbative `completion' has several merits. Firstly it seems natural from the string worldsheet point of view. Second, it enjoys other nice mathematical properties as listed in \cite{Benjamin:2023nts}. However, we would like to point out that contributions from the negative winding modes is \emph{not} the non-perturbative correction we have seen from resurgence analysis for several reasons.\par

From the spectrum, it is easy to see that $\mathcal{E}_n(\lambda)$ and $\mathcal{E}_n^{\text{np}}(\lambda)$ are solutions of \emph{different} Burgers' equations. In fact, $\mathcal{E}^{\text{np}}(\lambda)$ can be seen as the `negative branch' of the $T\overline{T}$-deformed theory with initial spectrum $(-E_n,\pm P_n)$ with deformation parameter $-\lambda$. Therefore, starting with the theory with spectrum $(E_n,P_n)$ and perform the $T\overline{T}$-deformation, the only deformed spectrum are $\mathcal{E}^{\pm}_n(\lambda)$, and states with deformed energy $\mathcal{E}_n^{\text{np}}(\lambda)$ cannot be generated by the same $T\overline{T}$-deformation but has to be put in by hand. On the other hand, the resurgence analysis is solely based on the $T\overline{T}$-deformed theory with spectrum $(E_n,P_n)$, without extra input. Therefore it is unlikely that the negative winding mode will emerge in this way. For the full partition function, the non-perturbative contributions from $\mathcal{E}^-_n$ are absent for $\lambda>0$, and start to play a role at $\lambda<0$, according to the resurgence analysis, while the contributions from $\mathcal{E}_n^{\text{np}}$ is also present for positive $\lambda$, if negative winding modes are included.

\section{Discussions}
\label{sec:disucss}

In this paper, we settle two important problems concerning the $T\overline{T}$-deformation of conformal field theories, both for free or interacting rational CFTs. First of all, we solve the convergence problem of the perturbative expansion of the $T\overline{T}$-deformed torus partition function, and find that it is asymptotic, in the sense that the coefficients grow factorially fast. We benefit from a streamlined and extremely efficient algorithm to calculate the perturbative coefficients, exploiting both the recursion relations as well as the modular properties satisfied by them. Secondly, we find definitive evidence that the $T\overline{T}$-deformed torus partition function does receive non-perturbative corrections, albeit in the regime where the $T\overline{T}$ parameter $\lambda$ is negative. 
These non-perturbative corrections are found by analysing the asymptotic behavior of the perturbative coefficients using the resurgence theory, and they are matched by a careful saddle point analysis of the integral form of the $T\overline{T}$-deformed partition function. These non-perturbative corrections do not look exactly the same as the ones typically found in local QFTs, manifesting non-locality of the $T\overline{T}$ deformation. We also explain the appearance of the non-perturbative corrections in the negative $\lambda$ regime, but not in the positive $\lambda$ regime, through the Stokes' phenomenon.

It would be interesting to perform a similar analysis for other solvable irrelevant deformation such as the $J\bar{T}$-deformation \cite{Guica:2017lia} and the more general $T\bar{T}+J\bar{T}+\bar{J}T$-deformation. In particular, for $J\bar{T}$-deformation, the flow equation \cite{Aharony:2018ics} and the integral representation \cite{Hashimoto:2019wct} are known. An intriguing observation from $J\bar{T}$ deformation is that the series expansion actually truncates for the deformed free boson. It is therefore interesting to see whether similar truncations occur for free fermion and other interacting theories.\par

Another interesting direction is considering other physical quantities, such as the torus one-point function. For CFT, the torus partition one-point function of a primary operator is modular covariant. The modular properties of the deformed torus one-point function has been studied in \cite{Cardy:2022mhn}. Expanding the one-point function as a $\lambda$-series, the first few perturbative results have been computed \cite{He:2020udl} using conformal perturbation theory. It would be interesting to exploit the modular properties and compute high order perturbative results. From the resurgence analysis, we can extract possible non-perturbative contributions for the operator, which will be important for understanding non-perturbative corrections for general deformed correlation functions.\par

In this paper, we have compared the non-perturbative result from resurgence with the negative winding modes for $w=-1$. It would be desirable to perform a more systematic resurgence analysis for the single-trace  $T\bar{T}$-deformed partition function, based on the methods developed in this paper. The interesting new feature is taking into account the contributions from twisted sectors. Once this is achieved, a concrete goal is performing a detailed resurgence analysis for the Sym$^N(\mathbb{T}^4)$ theory, which is dual to the tensionless string in AdS$_3\times S^3\times\mathbb{T}^4$ with $k=1$. The identification of non-perturbative contributions in the bulk would be very interesting.\par

The structure of the $T\overline{T}$ coefficients $Z_n$, as exemplified in \eqref{eq:Znchil}, is quite universal, at least as far as free theories and rational CFTs are concerned. In particular, the key formulae \eqref{eq:recurschi2}, \eqref{eq:DD-fermion}, \eqref{eq:DD-LY} for calculating the coefficients in \eqref{eq:Znchil} only depend on the modular weight of the undeformed and holomorphic character $\chi_0$. 
It is the same for the free fermion and the Lee-Yang model, but different for the free boson. The reason is that the non-compact free boson we considered in this paper has a non-trivial zero mode. Incidentally, the zero mode is also responsible for different values of $\nu$, as in \eqref{eq:pars-nu}, between the free boson model and the free fermion as well as the Lee-Yang model, as explained in Section~\ref{sec:Saddles}. Therefore, we expect that among rational CFTs all the minimal models should share not only the same structure \eqref{eq:Znchil} for the $T\overline{T}$ coefficients $Z_n$, but also the same values of the coefficients $c_{n,\ell}$, as well as the identical $\nu$ value. It would be interesting to see this explicitly in more examples of rational CFTs, as well as to extend the analysis to close cousins like supersymmetric rational CFTs. \par

While we expect our results to be representative of rational CFTs, a tempting next step is to extend the analysis to include more general CFTs, particularly the chaotic ones. In our analysis, an important strategy in identifying the relevant saddle-points is to focus on the singular behaviors of the undeformed CFTs partition function. Though such behaviors in chaotic CFTs are rarely presented through exact expressions like those of rational CFTs, they often emerge via the phenomenon of universality, which chaotic systems are usually believed to be blessed with. For example, the spectral form factor of chaotic systems, which is closely related to the partition function with analytically continued modular parameters, has been studied extensively and shown to exhibit universal behaviors \cite{Cotler_2017,saad2019semiclassicalrampsykgravity,Dyer_2017,Benjamin_2019,Kudler_Flam_2020,10.21468/SciPostPhys.8.6.088}. In the future it is interesting to explore how our analysis can be adopted to include such universal properties of chaotic CFTs.    

In this paper we focused solely on the limit of small $\lambda$. Equipped with an improved understanding of its non-perturbative effects, it would be interesting to study how our analysis is affected by taking other physical limits first. For example, we can study the high temperature limit $\tau_2\to 0$, that requires crossing the Hagedorn-like singularity at $\tau^c_2$ for $\lambda>0$. As was alluded to, we possibly need to re-sum the corrections to the integrand of (\ref{eq:integral_rep}) from all the winding-mode contributions, and then perform the integral of the dynamical modular parameters $\lbrace\zeta,\bar{\zeta}\rbrace$. A plausible guess is that the subsequent saddle-point analysis in $\lambda$ could exhibit a change in nature when the physical temperature $\tau_2$ varies across $\tau^c_2$. We will pursue this difficult yet important direction in the future. We can also generalize our analysis to complex $\tau$ by turning on an angular potential $\tau_1$, after which we can examine the limit of $\tau \to 0$ along general complex directions. Another important class is the holographic limit $c\to \infty$. Taking this before the small $\lambda$ limit amounts to approximating $Z_{\text{CFT}}(\zeta,\bar{\zeta})$ in (\ref{eq:integral_rep}) by the gravity result, which in itself exhibits very rich non-perturbative phenomenon in small $1/c$ or equivalently small $G_N$ in the bulk. Studying how they interplay with the saddle-point analysis in small $\lambda$ is also an interesting investigation for the future.   

To handle asymptotic series beyond Gevrey-1, standard tool boxes in resurgence analysis should also be extended. For example, the usual Richardson transformation which one applies to extract stable values does not apply here and a generalized version is needed. Also, the asymptotic $\lambda$-series is not Borel resummable, which hinders us from obtaining closed form results from perturbative data. Is there a method which allows us to do so ? Such questions are interesting for resurgence theory and we leave them for future works.


\section*{Acknowledgment}
We thank Andrea Dei and Wei Song for helpful discussions on holography.
The work of J.G. is partly supported by Startup Funding no.~4007022316 and 4007022411 of Southeast University and by the NSF of China through Grant No.~6507024099.
The work of Y.J. is partly supported by Startup Funding no. 3207022217A1 of Southeast University and by the NSF of China through Grant No. 12247103.
The work of H.W. is supported by the NSF of China through Grant No. 12175238.

\appendix

\section{Generalised Richardson transform}
\label{sec:Richardson}

Given the asymptotic behavior of a function $f(n)$ at large $n$ 
\begin{equation}
    f(n) \sim f_0 + \frac{f_k}{n^k} +\mc{O}(n^{-k-1}),\quad n\rightarrow \infty
\end{equation}
with some $k\geq 1$,
the standard method to remove the $1/n^k$ term, even if $f_k$ is not known, so that we approach $f_0$ faster as $n\rightarrow \infty$, is the Richardson transform of the $k$-th order
\begin{equation}
\label{eq:Rkf}
    R_k[f](n) = \frac{n^kf(n)-(n-s)^kf(n-s)}{n^k - (n-s)^k},
\end{equation}
where $s$ is any integer greater than zero,
as 
\begin{equation}
    R_k[f](n) \sim f_0 + \mc{O}(n^{-k-1}),\quad n\rightarrow \infty.
\end{equation}
See for instance \cite{Bender78}.

One can actually show that this works for any $k\in \IQ$.
For instance, if 
\begin{equation}
    f(n) \sim f_0 + \frac{f_{1/2}}{n^{1/2}} + \mc{O}(1/n),\quad n\rightarrow \infty,
\end{equation}
then
\begin{equation}
    R_{1/2}[f](n)\sim f_0 +\mc{O}(1/n),\quad n\rightarrow\infty,
\end{equation}
where $R_k[f](n)$ is \eqref{eq:Rkf} with $k=1/2$.
We will call this the generalised Richardson transform.

\section{More saddle-points from full modular invariance}
\label{sec:modular_images}

For the actual CFT partition function, it enjoys the full modular transformation properties: 
\be
Z\left(\tau',\bar{\tau}'\right) = Z(\tau,\bar{\tau}),\;\;\tau' = \frac{a\tau+b}{e\tau+d},\;\;\bar{\tau}' = \frac{a\bar{\tau}+b}{e\bar{\tau}+d}  
\ee
Due to this, the singularities of the actual CFT partition function is much richer than that of the toy model (\ref{eq:toy_model}). The singularity $\tau\to 0,\;\bar{\tau}\to \infty$ captured in (\ref{eq:toy_model}) is not the only divergence of the partition function. Their images under modular transformations are also singularities of the CFT patition function. For rational CFTs, the leading-order singular behavior is preserved also under:
\be\label{eq:independent_modular} 
\tau \to \frac{a\tau+b}{e\tau+d},\;\;\bar{\tau} \to \frac{\alpha \bar{\tau}+\beta}{\gamma \bar{\tau}+\delta} 
\ee
which amounts to two independent modular transformations for $\tau$ and $\bar{\tau}$ respectively. From this we can infer that rational CFT partition functions feature additional singularities that include also the images under the more general transformations (\ref{eq:independent_modular}). 

The mechanism of balancing the $T\overline{T}$ kernel against diverging CFT free energy also applies near these additional singularities, and could give rise to additional saddle-points. We can analyze a generic one of them by considering the following toy-model integral:
\bea
\mathcal{Z}(\tau_2|\lambda) &=& \int_{\mathcal{H}^+} \frac{\rd^2 \zeta}{\zeta^2_2} \exp\left[\frac{2\tau_2}{\lambda}-\frac{2\ri(\zeta \bar{\zeta}+\tau_2^2)}{\lambda(\zeta-\bar{\zeta})}\right]\nonumber\\
&\times & \exp\left[-\frac{\pi c \ri}{12}\left(\frac{a \zeta+b}{e\zeta+d}-\frac{\alpha \bar{\zeta}+\beta}{\gamma \bar{\zeta}+\delta}\right)\right] 
\eea
Like what (\ref{eq:toy_model}) does, the second-line is supposed to approximate the seed-CFT partition function near: 
\begin{eqnarray}\label{eq:modular_singular}
\zeta = -d/e,\;\;\;\bar{\zeta} = -\delta/\gamma    
\end{eqnarray}
which are the modular transformed images of the singularities in (\ref{eq:toy_model}). For simplicity we assume that they are neither zero nor infinity. The saddle-point equations still admit perturbative solutions in $\lambda^{1/2}$ near (\ref{eq:modular_singular}):
\bea
\zeta^* =-\frac{d}{e}\pm \sqrt{\frac{\pi c \lambda}{24}}\frac{(d\gamma-e\delta)}{e^2\sqrt{\delta^2+\gamma^2 \tau_2^2}}+... \nonumber\\
\bar{\zeta}^* = -\frac{\delta}{\gamma} \pm \sqrt{\frac{\pi c \lambda}{24}}\frac{(d\gamma-e\delta)}{\gamma^2\sqrt{d^2+e^2 \tau_2^2}}+... 
\eea
Plugging the saddle-point solution back into the integral, we obtain the leading order contributions: 
\bea\label{eq:more_saddle}
\mathcal{Z}(\tau_2,\lambda) &\sim & \exp\left[\lambda^{-1}\left(2\tau_2 \pm 2\ri  \Delta\right)+...\right]\nonumber\\
\Delta &=& \frac{d\delta+e\gamma \tau_2^2}{d\gamma-e\delta}
\eea
We see that at the additional saddle-points, the leading order action receives an imaginary part proportional to $\Delta$. It can then be deduced that the corresponding contribution $Z^\Delta_n$ to the asymptotic behavior of the expansion coefficients takes the form:
\bea
Z^\Delta_n &\sim & n!\tilde{A}^{-n} \cos{\left(k n\right)},\;\;k = \tan^{-1}(-\Delta/\tau_2) \nonumber\\
\tilde{A} &=& 2\sqrt{\tau_2^2+\Delta^2} \geq A = 2\tau_2
\eea 
Most prominently, it contains an oscillatory factor and decays exponentially faster than the saddle-points we found in the main text. In principle, there could be infinitely many of them. Their combined effects may explain the interesting behavior we observe in the actual $T\overline{T}$-deformed CFT partition functions. 

\section{Stokes' phenomenon: a quick review}
\label{sec:StokesReview}
In this appendix, we recall the basic ingredients underlying the Stokes' phenomenon. A prototypical setting involves the integral of the form: 
\begin{equation}\label{eq:Stokes_example}
F(k) = \int_{\gamma} \rd x\; e^{-k I(x)}
\end{equation}
where $k$ is in general a complex parameter with large modulus. In order for the integral to be well-defined, the integration contour $\gamma$ has to asymptote towards the converging region where $\text{Re}\; k I(x) \to \infty$. The integral can be approximated by summing over contributions from the saddle-points $x_m$ satisfying: $I'(x_m) = 0$. In particular, for any given saddle one can construct the steepest descent contour $\mathcal{J}_m$ passing through $x_m$, known as the Lefschetz thimble. For one-dimensional integral $\mathcal{J}_m$ is characterized by the conditions: 
\bea\label{eq:Lefschetz_def}
\text{Im}\; k I(x) = \text{Im}\; k I(x_m),\;\; \text{Re}\; k I(x) \geq \text{Re}\; k I(x_m)
\eea
for $x\in \mathcal{J}_m$. For higher dimensional integrals, the Lefschetz thimbles are defined as being generated by the gradient flows of the real part $h(x)=\text{Re}\;k I(x)$ \cite{fedoryuk:1977saddle}. In general, for a $n$ real-dimensional integral defined in $n$ complex-dimensions, the Lefschetz thimbles $\mathcal{J}_m$ are $n$ real-dimensional submanifolds of $\mathbb{C}^n$ through $x_m$. The conditions (\ref{eq:Lefschetz_def}) are still satisfied on $\mathcal{J}_m$, but are not sufficient to define $\mathcal{J}_m$ \cite{ursell:1980integrals,kaminski:1994exponentially}. The contribution from a particular saddle-point $x_m$ can therefore be written canonically as: 
\be
F_m(k) = \int_{\mathcal{J}_m} \rd x\; e^{-k I(x)}
\ee
In the large $k$ limit, it admits an asymptotic expansion
\begin{equation}
    F_m(k) \sim \frac{e^{-k I(x_m)}}{\sqrt{kI''(x_m)/2}} \left(1+\cdots\right)
\end{equation}
where ... denotes higher order corrections about this saddle-point.
For a given saddle-point $x_m$, whether it contributes to the integral (\ref{eq:Stokes_example}) depends on the integration contour $\gamma$. In generally it can be decomposed into a union of the Lefschetz thimbles \cite{pham:1967introduction}: 
\be
\gamma = \cup_m c_m \mathcal{J}_m
\ee
where the coefficients $c_m$ take values in $(0,\pm 1)$ depending on whether it appears in the decomposition and its relative orientation with respect to $\gamma$. The integral can now be written formally as a sum of the saddle-point contributions: 
\be\label{eq:saddle_point_sum}
F(k) = \sum_m \; c_m F_m(k) \sim \sum_{m} c_m\;  \frac{e^{-k I(x_m)}}{\sqrt{kI''(x_m)/2}} \left(1+\cdots\right)
\ee
where $\sim$ denotes asymptotic expansion in $1/k$ near each saddle point. The full sum over these series through (\ref{eq:saddle_point_sum}) is called the trans-series. 

We emphasize that the definition of the Lefschetz thimbles depends on the coupling constant $k$, in particular on its phase $k = e^{\ri\theta}|k|$. Because of this, as $k$ rotates in the complex plane, the Lefschetz thimbles deform accordingly as submanifolds. The decomposition coefficients $c_m$ also changes, but only by integers when the Lefschetz thimble undergoes a topological change. There will be critical values for the phase $\theta$ across which the coefficients jump discontinuously, \emph{e.g.} $c_m = c_m \pm 1$. As a result, the saddle-point contributions to the integral (at fixed integration contour $\gamma$) also jump according to (\ref{eq:saddle_point_sum}). This is called the Stokes' phenomena.  The critical phase $\theta_m$ across which $c_m$ jumps describes a ray in the complex $k$-plane, that marks an edge of the wedge inside which the saddle-point contribution $F_m$ appears in (\ref{eq:saddle_point_sum}). It is the Stokes' ray associated with the saddle-point $x_m$. 

Without loss of generality, we can take the integration contour $\gamma$ to coincide with a particular Lefschetz thimble $\mathcal{J}_0$ through the saddle point $x_0$ at say $k=|k|>0$, and consider the Stokes' phenomenon associated with another saddle $x_m$. For one-dimensional integral, it can be described geometrically in terms of when the Lefschetz thimble $\mathcal{J}_0$ passes through $z_m$, so that both $z_m$ and $z_0$ lie on $\mathcal{J}_0$. When this happens $\mathcal{J}_0$ usually takes a sharp turn at $z_m$.
It is easy to see that this happens when: 
\begin{equation}\label{eq:Stokes_pheno}
\text{Im}\;k I(x_0) = \text{Im}\; k I(x_m),\;\;\; \text{Re}\;k I(x_0)< \text{Re}\;k I(x_m)
\end{equation}
Solutions to (\ref{eq:Stokes_pheno}) specifies a set of phases $\theta_m$ for $k$, \emph{i.e.} rays of $k$ -- they are the Stokes' rays. When $k$ crosses a Stokes' ray, say, associated to $z_m$, 
$\mathcal{J}_0$ undergoes a change in topology
\begin{equation}
    \mathcal{J}_0 \rightarrow \mathcal{J}_0 + \mathcal{J}_m,
\end{equation}
which implies that the value of $F_m(k)$ has a discontinuity
\begin{equation}
    \label{eq:DiscF}
    \text{Disc}_{\theta_m}F_0(k) := F_0(k^+) - F_0(k^-) = F_m(k), 
\end{equation}
with $k^{\pm} = |k|\re^{\ri(\theta_m\pm0^+)}$.

For higher dimensional integrals, the geometric description of the Stokes' phenomenon becomes obscure. However, it can be shown that the Stokes' phenomenon remains a real co-dimension one event, \emph{i.e.} it can be generically encountered by tuning only one real-parameter, e.g. the phase $\theta$ of $k$ \cite{howls:1997hyper}. When this happens, (\ref{eq:Stokes_pheno}) must still be satisfied. Technically, for higher dimensional integrals (\ref{eq:Stokes_pheno}) is only a necessary condition -- there are additional uncertainties due to the possible non-trivial topology of the Lefschetz thimbles, \emph{e.g.} they may contain multiple Riemann-sheets. 

\section{Resurgent relations}
\label{sec:ResRel}

\begin{figure}
    \centering
   \subfloat[Residue contour]{\includegraphics[width=0.6\linewidth]
   {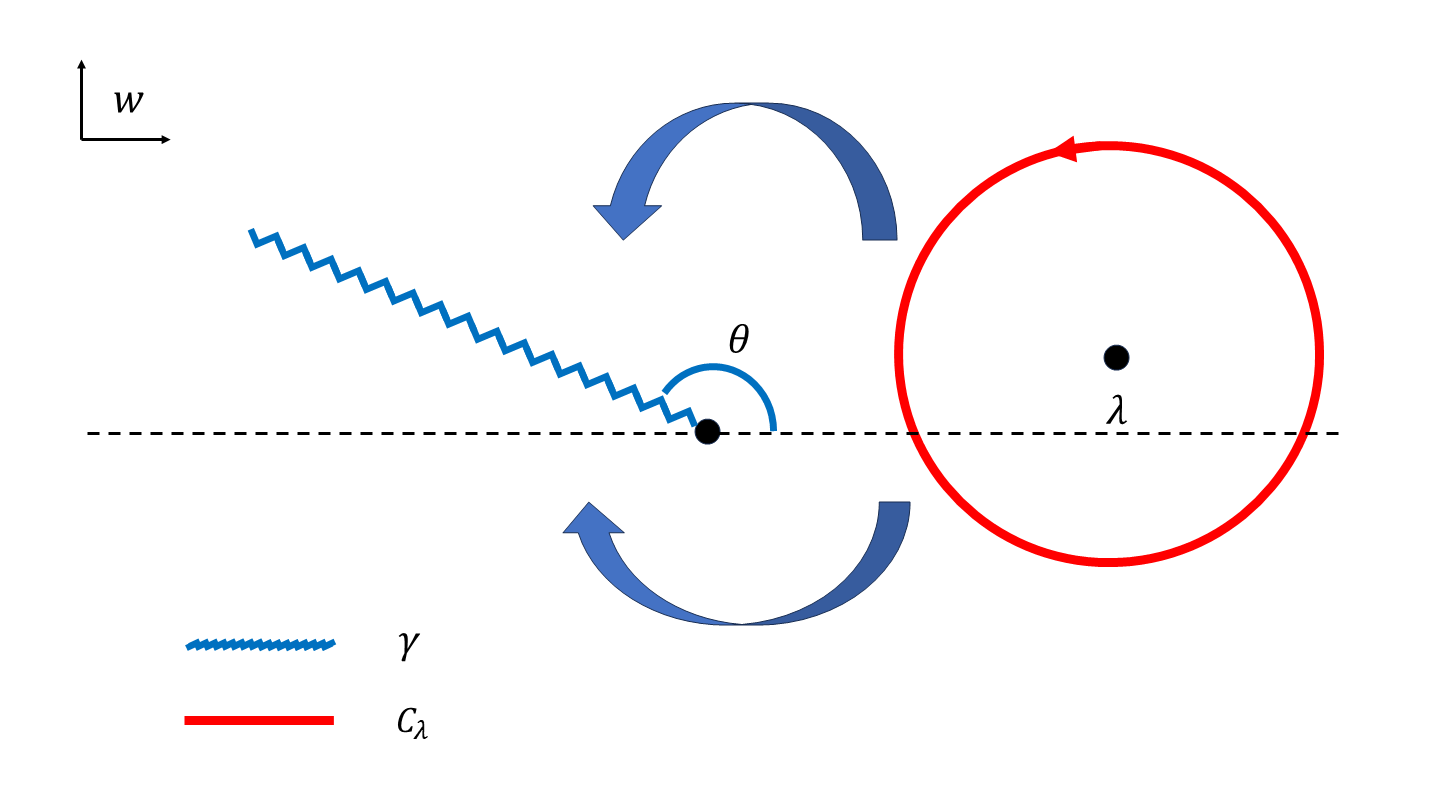}}\\
   \subfloat[and its deformation]{\includegraphics[width=0.6\linewidth]{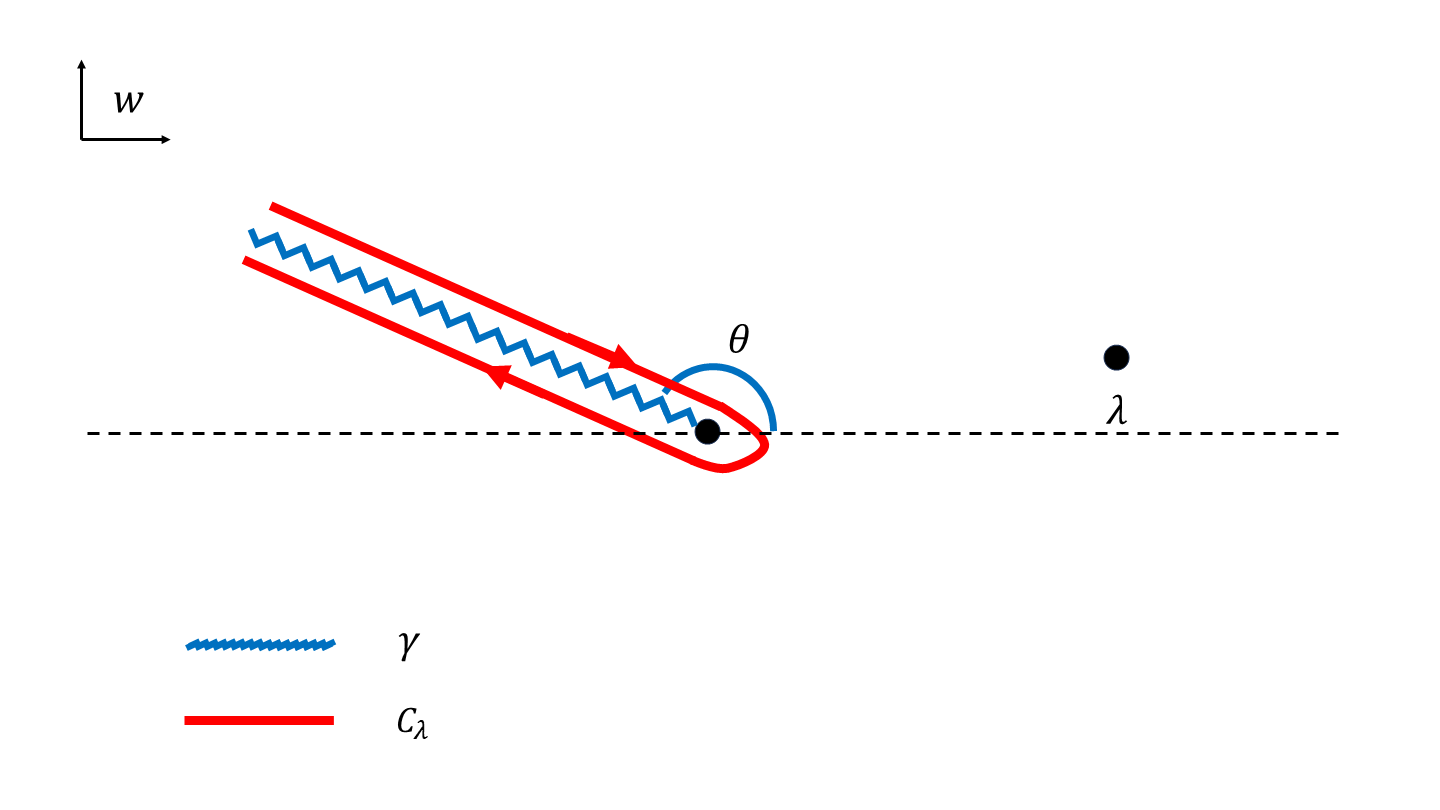}}
    \caption{Illustrated contour deformation for Eq (\ref{eq:contour_deform}). }
    \label{fig:contour_deform}
\end{figure}

Here we derive the resurgent relation \eqref{eq:b1A}, \emph{i.e.}, how a nonperturbative component of the type \eqref{eq:Znp} in the partition function affect the asymptotic behavior of the perturbative coefficients $Z_n$. Typically, factorial growth of perturbative coefficients mean the perturbative component $\mathcal{Z}^{\text{pert}}$ has discontinuity across certain ray $\rho_\theta = \IR \re^{\ri\theta}$, and the discontinuity is given by the nonperturbative component $Z^{\text{np}}$ associated to the ray
\begin{align}\label{eq:discZ}
    &\mathcal{Z}^{\text{pert}}(\lambda^+) - \mathcal{Z}^{\text{pert}}(\lambda^-)  = \text{Disc}_\theta\; \mathcal{Z}^{\text{pert}}(\lambda)
    = \mathcal{Z}^{\text{np}}(\lambda) \nonumber\\
    \sim 
    &\lambda^{-\nu} e^{-A/\lambda-B/\sqrt{\lambda}} \sum^{\infty}_{k=0} b_k \lambda^{k/2}
\end{align}
with $\arg\lambda^{\pm} =\theta\pm 0^+$.
Naturally, there could be multiple Stokes curves, associated with distinct non-perturbative corrections. They are branch-cuts for the perturbative component of the partition function. To proceed, we begin by writing the perturbative partition function $\mathcal{Z}^{\text{pert}}(\lambda)
\equiv \mathcal{Z}^{\text{pert}}(\tau,\bar{\tau}|\lambda)$ in terms of a contour integral and perform the following steps: 
\begin{align}
\label{eq:contour_deform}
    \mathcal{Z}^{\text{pert}}(\lambda) 
    &= \frac{1}{2\pi \ri}\oint_{\mathcal{C}_\lambda} \rd w\frac{\mathcal{Z}^{\text{pert}}(w)}{w-\lambda} \\\nonumber
    &= \frac{1}{2\pi \ri}\int_{\rho_\theta} \rd w \frac{\text{Disc}_\theta\;Z^{\text{pert}}(w)}{w-\lambda} 
\end{align}
where we first deformed the contour $\mathcal{C}_\lambda$ around $\lambda$ to wrap around the branch-cut $\rho_\theta$, making it an integral along $\rho_\theta$ of the discontinuous jump $\text{Disc}_\theta \mathcal{Z}^{\text{pert}}(w)$, see Figure (\ref{fig:contour_deform}).

By performing small $\lambda$ asymptotic expansion on both sides of \eqref{eq:contour_deform}, this leads to an integral representation of the perturbative expansion coefficient $Z_n$: 
\begin{align}
\label{eq:coeff_integral}
    Z_n 
    &= \frac{1}{2\pi \ri}\int_{\rho_\theta}\rd w\; w^{-(n+1)}\; Z^{\text{np}}(w) \\\nonumber
    &\sim \frac{1}{2\pi \ri}\int^{\re^{\ri\theta}\infty}_0 \rd w w^{-(n+1+\nu)} e^{-A/w-B/\sqrt{w}}\sum^\infty_{k=0} b_k w^{k/2}
\end{align}
where $\theta$ denotes the phase of the branch-cut ray $\rho_\theta$ originating from the origin.  We can directly perform the integral for each term in the sum. Doing this yields the explicit expansion formulae: 
\begin{align}
    Z_n &\sim \sum^\infty_{k=0}b_k \int^{\re^{i\theta}\infty}_0 \rd w\;e^{-\frac{A}{w}-\frac{B}{\sqrt{w}}}\;w^{-n-1-\nu+\frac{k}{2}}\\\nonumber
    &\sim  \sum^\infty_{k=0}b_k\;\frac{\Gamma(2n+2\nu-k)}{2^{2n+2\nu-k-1}A^{n+\nu-\frac{k}{2}}}\;U\left(n+\nu-\frac{k}{2},\frac{1}{2},\frac{B^2}{4A}\right)
\end{align}
where $U$ is the Tricomi confluent hypergeometric function. A possibly more illuminating approach is to perform a saddle-point approximation of (\ref{eq:coeff_integral}) at large order, using $n$ as the large parameter. This can be done for each term in the sum over $k$ separately. To this end, it is convenient to introduce $w = x^2$,
and rewrite the integral as:
\begin{equation}
  J_n^{(k)} \sim  \int_0^{\re^{\ri\theta}\infty} \re^{V(x)}\rd x,\;\;
\end{equation}
with
\begin{equation}
    V(x) = -\frac{A}{x^2} -\frac{B}{x} -2 n_k \log x,\quad
  n_k = n+\nu-\frac{k}{2}+\frac{1}{2}.
\end{equation}
For the resurgence analysis of $Z_n$ at large order $n$, it suffices to compute only the leading order terms, \emph{e.g.} $k=0, 1$. Higher-order terms affect the higher-order in $n^{-1}$ corrections in the asymptotic behaviour of $Z_n$. The saddle point equation takes the form:
\begin{equation}
  V'(x) = \frac{2A}{x^3}+ \frac{B}{x^2}- \frac{2n_k}{x} = 0 
\end{equation}
and admits two solutions:
\begin{equation}\label{eq:large_order_saddle}
  x_{\pm} = \frac{B}{4n_k} \pm\sqrt{\frac{B^2}{16n_k^2}+\frac{A}{n_k}}.
\end{equation}
It turns out that the large order behaviour of $J^{(k)}_n$ is controlled by the saddle point $x_+$, because it is closer to the the integration contour $(0,\re^{\ri\theta}\infty)$. We can then deform the contour to pass through $x_+$ and expand near it: 
\begin{equation}
  x = x_{+} + c \delta x,\;\;\;c=\left(\frac{3A}{x_+^4}+\frac{B}{x_+^3}-\frac{n_k}{x_+^2}\right)^{-1/2}
\end{equation}
where the coefficient $c$ is chosen such that the quadratic term is a standard Gaussian: 
\be 
V(x) \approx V(x_+) - \delta x^2 + \mathcal{O}(\delta x^3)
\ee
Performing the Gaussian integrals in $\delta x$ order by order yields the following 
asymptotic formula:
\begin{align}
  J_n^{(k)} \sim  
  &\re^{\frac{B^2}{8A}}\sqrt{\frac{\pi}{3}}\;
  \frac{\Gamma\left(n+\nu-\frac{k}{2}\right)}{A^{n+\nu-\frac{k}{2}}}
  \re^{- B\sqrt{\frac{n+\nu-\frac{k}{2}}{A}}}\\\nonumber
  &\left(1 + \frac{12AB-B^3}{96A^{3/2}\sqrt{n+\nu-\frac{k}{2}}} + \mc{O}\left(1/n\right)\right)
\end{align}
where we have used the Gaussian integral formula:
\begin{equation}
  \label{eq:gaussian}
  \int_{-\infty}^{+\infty} \rd x\;\re^{-x^2} x^{2n} = \Gamma(n+1/2).
\end{equation}
The asymptotic formula for the expansion coefficient $Z_n$ can then be assembled into:
\begin{equation}
  Z_n \sim \sum_{k=0}^\infty b_k J_n^{(k)}.
\end{equation}
For our purpose, we choose to keep corrections only up to $1/\sqrt{n}$, this involves only $k=0,1$ and we arrive at the resurgent relation \eqref{eq:b1A}.

\bibliography{references.bib}



\end{document}